\documentclass[usenatbib]{mnras}
\usepackage[T1]{fontenc}
\usepackage[latin9]{inputenc}
\usepackage{amsmath}
\usepackage{array}
\usepackage{multirow}
\usepackage{graphicx}
\providecommand{\tabularnewline}{\\}

\begin{document}
\title{
Solar dynamo cycle variations with a rotational period}
\author[V.V.~Pipin]
{V.V.~Pipin \thanks{email: pip@iszf.irk.ru}\\
{Institute of Solar-Terrestrial Physics, Russian Academy of
Sciences, Irkutsk, 664033, Russia} }
\maketitle 
\begin{abstract}
Using the non-linear mean-field dynamo models we calculate the magnetic
cycle parameters, like the dynamo cycle period, the amplitude of the
total magnetic energy, and the Poynting flux luminosity from the surface
for the solar analogs with rotation periods of range from 1 to 30
days. We do simulations both for the kinematic and non-kinematic dynamo
models. The kinematic dynamo models, which take into account the non-linear
$\alpha$-effect and the loss of the magnetic flux due to magnetic
buoyancy, show a decrease of the magnetic cycle with the decrease
of the stellar rotation period. The stars with a rotational period
of less than 10 days show the non-stationary long-term variations of
the magnetic activity. The non-kinematic dynamo models take into account
the magnetic field feedback on the large-scale flow and heat transport
inside the convection zone. They show the non-monotonic variation
of the dynamo period with the rotation rate. The models for the rotational
periods fewer than 10 days show the non-stationary evolution with a
slight increase in the primary dynamo period with the increase of
the rotation rate. The non-kinematic models show the growth of the
dynamo generated magnetic flux with the increase of the rotation rate.
There is a dynamo saturation for the star rotating with a period of two
days and less. The saturation of the magnetic activity parameters
is accompanied by depression of the differential rotation. 
\end{abstract}

\section{Introduction}

The partially convective stars, like the Sun, often show cyclic
magnetic activity \citep{Baliunas1995,Olah2009,Olspert2018}. The
magnetic activity of the Sun and solar-type stars demonstrates the
large-scale organization of the active phenomena both in time (cycles)
and in space \citep{Donati2009,See2016}. It is widely accepted that
the nature of the global magnetic activity in solar-type stars stems
from the turbulent hydromagnetic dynamo operating in their convection
zones. \cite{Parker1955} suggested the basic dynamo scenario for
the Sun. This scenario suggests the cyclic transformation of the large-scale
poloidal magnetic field into the toroidal magnetic field and vice
versa by means of the differential rotation and the cyclonic convection
motions. Hence, the energy supply for the dynamo process comes from
the energy of the global rotation and the turbulent energy of the
convective motions. The partially-convective stars with higher rotation
rates show a higher level of magnetic activity \citep{Noyes1984,Baliunas1995,See2015}.

\cite{Noyes1984} estimated the magnetic cycle parameters in solar-type
stars following the properties of the eigenmodes of the mean-field
dynamo equations. They found that the linear theory predicts the growing
level of magnetic activity. Simultaneously, the magnetic cycle
period decreases with the increase of the rotation rate of a star.
{Observations show that the magnitude of the surface latitudinal
shear is almost independent of the angular velocity for the solar-type
stars} (\citealp{Saar2011,Reinhold2013,Lehtinen2016}). This was found
in the mean-field models as well \citep{Kitchatinov1999a}. Therefore,
from the point of view of the linear theory, the dynamo efficiency
of the differential rotation does not increase with the increase
of the Sun's rotation rate \citep{Kitchatinov2011b}. On the other
hand, the growing level of the magnetic activity with the increase
of the rotation rate can be explained by the growing effect of the
Coriolis force acting on the convective flows. In the stratified convection
zone, this force results in the so-called $\alpha$ -effect \citep{Krause1980}.
This may explain why the Rossby number $\mathrm{Ro}=\mathrm{P_{rot}}/\tau_{c}$
is often a good parameter to trace the level of the magnetic activity
and cycle's parameters in the stars with the partial convection zone
\citep{Noyes1984,Baliunas1995,Olspert2018}.

Our main goal is to extend our previous study (see, \citealp{Pipin2015,Pipin2016b})
for the higher rotation rates employing the non-kinematic dynamo model,
which was developed recently by \cite{Pipin2019c}. The model takes
into account the magnetic activity effect on the large-scale flow
and heat transport in the convection zone. The latter effect seems
to results in the so-called ``extended cycle'' of the solar torsional
oscillations \citep{Ulrich2001}. The rotational and magnetic anisotropy
of the convective heat transport is rather important for the large-scale
flow organization in the stellar convection zone (see, e.g., \citealp{Busse1983,Kitchatinov1999a,Simitev2009,Kapyla2011,Gastine2012,Warnecke2013b}).
The recent global convection simulations of \cite{Strugarek2017,Warnecke2018A,Guerrero2019}
show the variations of the large-scale flow organization with the
rotation rate. Interestingly, that models of \cite{Warnecke2018A}
show that variations of the large-scale flow organization are accompanied
by the non-monotonic variations of the dynamo period with an increase
of the rotation rate. This indeed was found by \cite{Lehtinen2016}
and \cite{Lehtinen2020} in observations of the young solar-type stars.
These works encourage us to try the non-kinematic mean-field dynamo
models for the range of the rotational periods from 1 to 30 days.
For the sake of simplicity, we use the reference state of the model
which reproduces the one cell per hemisphere case. This type of meridional
circulation structure is popular in solar dynamo studies \citep{Charbonneau2011}.
Also, we do not take into account the evolutionary changes of the
stellar structure accepting the solar interior model of the modern
Sun. The next chapter describes some details of our dynamo model.

\section{Model}

We use the non-kinematic dynamo model developed by \cite[ hereafter PK19]{Pipin2019c}.
The model is based on the mean-field induction equation \citep{Krause1980}:
\begin{equation}
\mathrm{\partial_{t}\overline{\mathbf{B}}=\boldsymbol{\nabla}\times\left(\boldsymbol{\mathcal{E}}+\mathbf{\overline{U}}\times\overline{\mathbf{B}}\right)},\label{eq:mfe-1}
\end{equation}
where the induction vector of the large-scale magnetic field, $\overline{\mathbf{B}}$,
is represented as the sum of the toroidal and poloidal components:
\[
\overline{\mathbf{B}}=\hat{\mathbf{\boldsymbol{\phi}}}B+\nabla\times\frac{A\hat{\mathbf{\boldsymbol{\phi}}}}{r\sin\theta},
\]
where $r$ is the radial distance, $\theta$ is the polar angle, $\hat{\mathbf{\boldsymbol{\phi}}}$
is the unit vector in the azimuthal direction. The mean electromotive
force $\boldsymbol{\mathcal{E}}$ describes the turbulent generation
effects, pumping, and diffusion: 
\begin{equation}
\mathcal{E}_{i}=\left(\alpha_{ij}+\gamma_{ij}\right)\overline{B}_{j}-\eta_{ijk}\nabla_{j}\overline{B}_{k}.\label{eq:EMF-1-1}
\end{equation}
where the symmetric tensor $\alpha_{ij}$ stands for the turbulent
generation of the large-scale magnetic field by kinetic and magnetic
helicities; the antisymmetric tensor $\gamma_{ij}$ describes the
turbulent pumping effect including the mean-field magnetic buoyancy
\citep{Kitchatinov1993}; the anisotropic (in the general case) tensor
$\eta_{ijk}$ is the eddy diffusivity \citep{Pipin2018b}.

We employ the $\alpha$-effect tensor, in the following form: 
\begin{eqnarray}
\alpha_{ij} & = & C_{\alpha}\psi_{\alpha}(\beta)\alpha_{ij}^{(H)}+\alpha_{ij}^{(M)}\psi_{\alpha}(\beta)\frac{\overline{\chi}\tau_{c}}{4\pi\overline{\rho}\ell_{c}^{2}},\label{alp2d}
\end{eqnarray}
{where $\tau_{c}$ and $\ell_{c}$ are the convective turnover
time and the convective mixing length, respectively,} $\alpha_{ij}^{(H)}$
is hydrodynamic part of the $\alpha$-effect tensor; $\overline{\chi}=\overline{\mathbf{a\cdot b}}$
is the magnetic helicity density ($\mathbf{a}$ and $\mathbf{b}$
are the turbulent parts of the magnetic vector potential and magnetic
field vector), and tensor $\alpha_{ij}^{(M)}$ takes into account
the effect of the Coriolis force. Function $\psi_{\alpha}(\beta)$
stands for the ``algebraic'' saturation of the $\alpha$- effect
caused by the small-scale Lorentz force which opposes convective motions
across the field lines of the large-scale magnetic field, where, $\mathrm{\beta=\left|\overline{\mathbf{B}}\right|/\sqrt{4\pi\overline{\rho}u_{c}^{2}}}$.
For the strong magnetic field, when $\beta\gg1$, $\psi_{\alpha}(\beta)\sim\beta^{-3}$.
A detailed description of $\alpha_{ij}^{(H)}$, $\alpha_{ij}^{(M)}$
and $\psi_{\alpha}(\beta)$ is given by \cite{Pipin2018b}. The magnetic
helicity evolution follows the conservation law: 
\begin{equation}
\frac{\partial\overline{\chi}^{(tot)}}{\partial t}=-\frac{\overline{\chi}}{R_{m}\tau_{c}}-2\eta\overline{\mathbf{B}}\cdot\mathbf{\overline{J}}-\boldsymbol{\nabla\cdot\mathbf{F}}-\mathbf{\left(\overline{U}\cdot\boldsymbol{\nabla}\right)}\overline{\chi}^{(tot)}\label{eq:helcon}
\end{equation}
{where the first term in the RHS reflects the contribution of the
small-scale current helicity, $-2\eta$$\overline{\mathbf{b\cdot j}}$,
(see, \citealp{Kleeorin1999}), also,} 
\begin{equation}
\overline{\chi}^{(tot)}=\overline{\mathbf{A}\cdot\mathbf{B}}=\overline{\mathbf{A}}\cdot\overline{\mathbf{B}}+\overline{\mathbf{a\cdot b}}\text{,}\label{chitot}
\end{equation}
where $\mathbf{B}=\nabla\times\mathbf{A}$, $\overline{\mathbf{A}}$
is the large-scale magnetic field vector potential; $R_{m}$ is the
magnetic Reynolds number, (we put $R_{m}=10^{6}$). We assume that
the eddy diffusivity of the magnetic helicity is isotropic and that
the diffusive helicity flux $\mathbf{F}=-\eta_{\chi}\boldsymbol{\nabla}\overline{\chi}$,
where $\eta_{\chi}=0.1\eta_{T}$ \citep{Mitra2010}.

The large-scale flow, $\mathbf{\overline{U}}=\mathbf{\overline{U}}^{m}+r\sin\theta\Omega\left(r,\theta\right)\hat{\mathbf{\boldsymbol{\phi}}}$
includes effect of the differential rotation, $\Omega\left(r,\theta\right)$
and the meridional circulation, $\mathbf{\overline{U}}^{m}$. These
parameters are governed by the angular momentum conservation: 
\begin{eqnarray}
\frac{\partial}{\partial t}\overline{\rho}r^{2}\sin^{2}\theta\Omega & = & -\boldsymbol{\nabla\cdot}\left(r\sin\theta\overline{\rho}\left(\hat{\mathbf{T}}_{\phi}+r\sin\theta\Omega\mathbf{\overline{U}^{m}}\right)\right)\label{eq:angm}\\
 & + & \boldsymbol{\nabla\cdot}\left(r\sin\theta\frac{\overline{\mathbf{B}}\overline{B}_{\phi}}{4\pi}\right),\nonumber 
\end{eqnarray}
and by equation for the azimuthal component of large-scale vorticity,
$\mathrm{\overline{\omega}=\left(\boldsymbol{\nabla}\times\overline{\mathbf{U}}^{m}\right)_{\phi}}$:

\begin{eqnarray}
\mathrm{\frac{\partial\omega}{\partial t}\!\!\!} & \mathrm{\!\!=\!\!\!\!} & \mathrm{r\sin\theta\boldsymbol{\nabla}\cdot\left(\frac{\hat{\boldsymbol{\phi}}\times\boldsymbol{\nabla\cdot}\overline{\rho}\hat{\mathbf{T}}}{r\overline{\rho}\sin\theta}-\frac{\mathbf{\overline{U}}^{m}\overline{\omega}}{r\sin\theta}\right)}\label{eq:vort}\\
 & + & \mathrm{r}\sin\theta\frac{\partial\Omega^{2}}{\partial z}-\mathrm{\frac{g}{c_{p}r}\frac{\partial\overline{s}}{\partial\theta}}\nonumber \\
 & + & \frac{1}{4\pi\overline{\rho}}\left(\overline{\mathbf{B}}\boldsymbol{\cdot\nabla}\right)\left(\boldsymbol{\nabla}\times\overline{\mathbf{B}}\right)_{\phi}-\frac{1}{4\pi\overline{\rho}}\left(\left(\boldsymbol{\nabla}\times\overline{\mathbf{B}}\right)\boldsymbol{\cdot\nabla}\right)\overline{\mathbf{B}}{}_{\phi},\nonumber 
\end{eqnarray}
where $\hat{\mathbf{T}}$ is the turbulent stress tensor: 
\begin{equation}
\hat{T}_{ij}=\left(\left\langle u_{i}u_{j}\right\rangle -\frac{1}{4\pi\overline{\rho}}\left(\left\langle b_{i}b_{j}\right\rangle -\frac{1}{2}\delta_{ij}\left\langle \mathbf{b}^{2}\right\rangle \right)\right),\label{eq:rei}
\end{equation}
(see detailed description in PK19). Also, $\overline{\rho}$ is the
mean density, $\mathrm{\overline{s}}$ is the mean entropy; $\mathrm{\partial/\partial z=\cos\theta\partial/\partial r-\sin\theta/r\cdot\partial/\partial\theta}$
is the gradient along the axis of rotation. {The second line
in the Eq(\ref{eq:vort}) accounts the source terms of the meridional
circulation. They are due to the centrifugal and baroclinic forces.
In our models, we neglect the effects of the rotational oblateness
of the density and pressure profiles.} The mean heat transport equation
determines the mean entropy variations from the reference state due
to the generation and dissipation of the large-scale magnetic field
and large-scale flows \citep{Pipin2000}: 
\begin{equation}
\overline{\rho}\overline{T}\left(\frac{\partial\overline{\mathrm{s}}}{\partial t}\!+\!\left(\!\overline{\mathbf{U}}\cdot\boldsymbol{\nabla}\right)\!\overline{\mathrm{s}}\!\right)\!=\!-\!\boldsymbol{\nabla}\!\cdot\!\left(\!\mathbf{F}^{c}\!+\!\mathbf{F}^{r}\!\right)\!-\!\hat{T}_{ij}\frac{\partial\overline{U}_{i}}{\partial r_{j}}\!-\!\boldsymbol{\boldsymbol{\mathcal{E}}}\!\cdot\!\left(\!\nabla\!\times\!\overline{\boldsymbol{B}}\!\right),\label{eq:heat}
\end{equation}
where $\overline{T}$ is the mean temperature, $\mathbf{F}^{r}$ is
the radiative heat flux, $\mathbf{F}^{c}$ is the anisotropic convective
flux. For the anisotropic convective flux, we employ the expression
suggested by \cite{Kitchatinov1994}: 
\begin{equation}
\mathrm{F_{i}^{c}=-\overline{\rho}\overline{T}\chi_{ij}\nabla_{j}\overline{s}.}\label{conv}
\end{equation}
Further details about the dependence of the eddy thermal conductivity
tensor, $\mathrm{\chi_{ij}}$, on the global rotation and large-scale
magnetic field are given in{ \cite{Pipin2019c}}. {Note,
that the mean-field expressions of the convective heat flux were studied
in the global convection simulations of \cite{Warnecke2013}. Their
results suggest that anisotropy of the convective heat flux is really
important. Note, that in the global convection simulations, the approximation
of the heat flux via the mean entropy gradient was found qualitatively
correct only for the radial convective heat flux.} {For calculation
of $\hat{\mathbf{T}}$, $\boldsymbol{\boldsymbol{\mathcal{E}}}$ and
$\mathbf{F}^{c}$, we employ analytical results that were obtained
earlier using the mean-field magnetohydrodynamics framework. These
results take into account the effects of the global rotation and magnetic
field on turbulence. The details can be found in \cite{Pipin2019c}}.
The last two terms in Eq (\ref{eq:heat}) take into account the convective
energy gain and sink caused by the dynamo action, as well as, generation
and dissipation of the large-scale flows. The kinetic coefficients
in the mean-field analytical expressions of the mean electromotive
force, $\boldsymbol{\mathcal{E}}$ , and the turbulent stress tensor
$\hat{\mathbf{T}}$, depend on profiles of the turbulent parameters
of the convection zone, such as the typical convective turnover time,
$\tau_{c}$, and the convective RMS velocity, $u_{c}$. The reference
profiles of mean thermodynamic parameters, such as entropy, density,
and temperature are determined from the stellar interior model MESA
\citep{Paxton2011,Paxton2013}. The convective RMS velocity is determined
from the mixing-length approximation, 
\begin{equation}
\mathrm{u_{c}=\frac{\ell_{c}}{2}\sqrt{-\frac{g}{2c_{p}}\frac{\partial\overline{s}}{\partial r}},}\label{eq:uc}
\end{equation}
where $\ell_{c}=\alpha_{MLT}H_{p}$ is the mixing length, $\alpha_{MLT}=1.9$
is the mixing length parameter, and $H_{p}$ is the pressure height
scale. We determine the convective turnover time $\tau_{c}=\ell_{c}/\mathrm{u}_{c}$
from the parameters of the MESA code output. We assume that $\tau_{c}$
does not depend on the magnetic field and global flows. Eq.~(\ref{eq:uc})
determines the reference profiles for the eddy heat conductivity,
$\chi_{T}$, eddy viscosity, $\nu_{T}$, and eddy diffusivity, $\eta_{T}$,
as follows, 
\begin{eqnarray}
\chi_{T} & = & \frac{\ell^{2}}{6}\sqrt{-\frac{g}{2c_{p}}\frac{\partial\overline{s}}{\partial r}},\label{eq:ch}\\
\nu_{T} & = & \mathrm{Pr}_{T}\chi_{T},\label{eq:nu}\\
\eta_{T} & = & \mathrm{Pm_{T}\nu_{T}}.\label{eq:et}
\end{eqnarray}
The model gives the best agreement of the angular velocity profile
with helioseismology results for $\mathrm{Pr}_{T}=3/4$ (PK19). Also,
the dynamo model reproduces the solar magnetic cycle period, $\sim22$
years, if $\mathrm{Pm}_{T}=10$. For the solar case, we use the period
of rotation of solar tachocline determined from helioseismology, $\Omega_{0}/2\pi$=430
nHz \citep{Kosovichev1997}, which corresponds to the sidereal period
of about $\mathrm{P_{rot}}=25$ days.

\subsection{Boundary conditions and tachocline}

The position of the top boundary is $r_{\mathrm{top}}=0.99R_{\odot}$,
the bottom of the convection zone is fixed to $r_{b}=0.728R_{\odot}$
, and the bottom of the tachocline is $r_{\mathrm{ta}}=0.68R_{\odot}$.
At the $r=r_{\mathrm{ta}}$ we put the solid body rotation and the
perfect conductor boundary conditions. We do not solve the heat transport
equation for the tachocline region. Instead, we assume that within
tachocline all turbulent coefficients (except the eddy viscosity and
eddy diffusivity) decrease factor of $\exp\left(-100z/R\right)$,
where $z$ is the distance from the bottom of the convection zone.
We restrict the decrease of the eddy viscosity and eddy diffusivity
by one order of magnitude for the numerical stability. At the top,
$r=r_{\mathrm{top}}$we employ the stress-free boundary condition
for the angular momentum problem. For the heat transport at the bottom
of the convection zone, $r_{b}=0.728R_{\odot}$ , we put the total
flux $\mathrm{F_{r}^{conv}+F_{r}^{rad}={\displaystyle \frac{L_{\star}\left(r_{b}\right)}{4\pi r_{b}^{2}}}}$
and for the external boundary, in following to \cite{Kitchatinov2011b}
, we use 
\begin{equation}
\mathrm{F_{r}=\frac{L_{\star}}{4\pi r_{top}^{2}}\left(1+\left(\frac{\overline{s}}{c_{p}}\right)\right)^{4}.}\label{eq:fr}
\end{equation}
The relative variations of the radiation flux are 
\begin{equation}
\frac{\delta F_{r}}{F_{\odot}}=\left(1+\left(\frac{\overline{s}}{c_{p}}\right)\right)^{4}.\label{eq:dfr}
\end{equation}

Following ideas of \cite{Moss1992} and \cite{Pipin2011a}, we formulate
the top boundary condition in the form that allows penetration of
the toroidal magnetic field to the surface: 
\begin{eqnarray}
\delta\frac{\eta_{T}}{r_{\mathrm{top}}}B\left(1+\left(\frac{\left|B\right|}{B_{\mathrm{esq}}}\right)\right)+\left(1-\delta\right)\mathcal{E}_{\theta} & = & 0,\label{eq:tor-vac}
\end{eqnarray}
where $r_{\mathrm{top}}=0.99R_{\odot}$, and parameter $\delta=0.99$
and $B_{\mathrm{esq}}=50$G. The magnetic field potential outside
the domain is 
\begin{equation}
A^{(vac)}\left(r,\mu\right)=\sum a_{n}\left(\frac{r_{\mathrm{top}}}{r}\right)^{n}\sqrt{1-\mu^{2}}P_{n}^{1}\left(\mu\right),\label{eq:vac-dec}
\end{equation}
where $\mu=\cos\theta$. The boundary conditions Eq(\ref{eq:tor-vac})
provide the Poynting flux luminosity of the magnetic energy out of
the convection zone: 
\begin{eqnarray}
\mathrm{L_{P}} & = & -\frac{1}{2}\int_{-1}^{1}\mathcal{E}_{\theta}\overline{B}_{\phi}\mathrm{d}\mu\label{eq:fcfb}\\
 & = & \frac{1}{2}\left(\frac{\delta}{1-\delta}\right)\frac{\eta_{T}}{r_{\mathrm{top}}}\int_{-1}^{1}B^{2}\left(1+\left(\frac{\left|B\right|}{B_{\mathrm{esq}}}\right)\right)\mathrm{d}\mu.\nonumber 
\end{eqnarray}
Also, we will consider the following integral parameters of the models,
the total toroidal magnetic flux in the convection zone:

\[
F_{T}=2\pi\int_{-1}^{1}\int_{r_{b}}^{r_{\mathrm{top}}}\left|\overline{B}_{\phi}\right|\sin\theta r^{2}\mathrm{d}r\mathrm{d}\mu,
\]
the total toroidal magnetic flux in subsurface layer 
\[
F_{S}=2\pi\int_{-1}^{1}\int_{r_{s}}^{r_{\mathrm{top}}}\left|\overline{B}_{\phi}\right|\sin\theta r^{2}\mathrm{d}r\mathrm{d}\mu,
\]
where $r_{s}=0.89R$.

We define the parameters characterizing the energy of the symmetric
and antisymmetric parts of the subsurface toroidal magnetic field:

\begin{eqnarray*}
\overline{E}_{B}^{S} & =\frac{1}{4} & \int_{-1}^{1}\left[\overline{B}_{\phi}\left(\mu,t\right)+\overline{B}_{\phi}\left(-\mu,t\right)\right]^{2}d\mu,\\
\overline{E}_{B}^{N} & =\frac{1}{4} & \int_{-1}^{1}\left[\overline{B}_{\phi}\left(\mu,t\right)-\overline{B}_{\phi}\left(-\mu,t\right)\right]^{2}d\mu.
\end{eqnarray*}
Then, the parity index, or the reflection symmetry index for this
component of the magnetic activity is

\begin{equation}
P_{B}=\frac{\overline{E}_{B}^{S}-\overline{E}_{B}^{N}}{\overline{E}_{B}^{S}+\overline{E}_{B}^{N}}.\label{eq:parity}
\end{equation}

\subsection{Turbulence parameters and reference models of the large-scale flow}

\begin{figure}
\includegraphics[width=0.99\columnwidth]{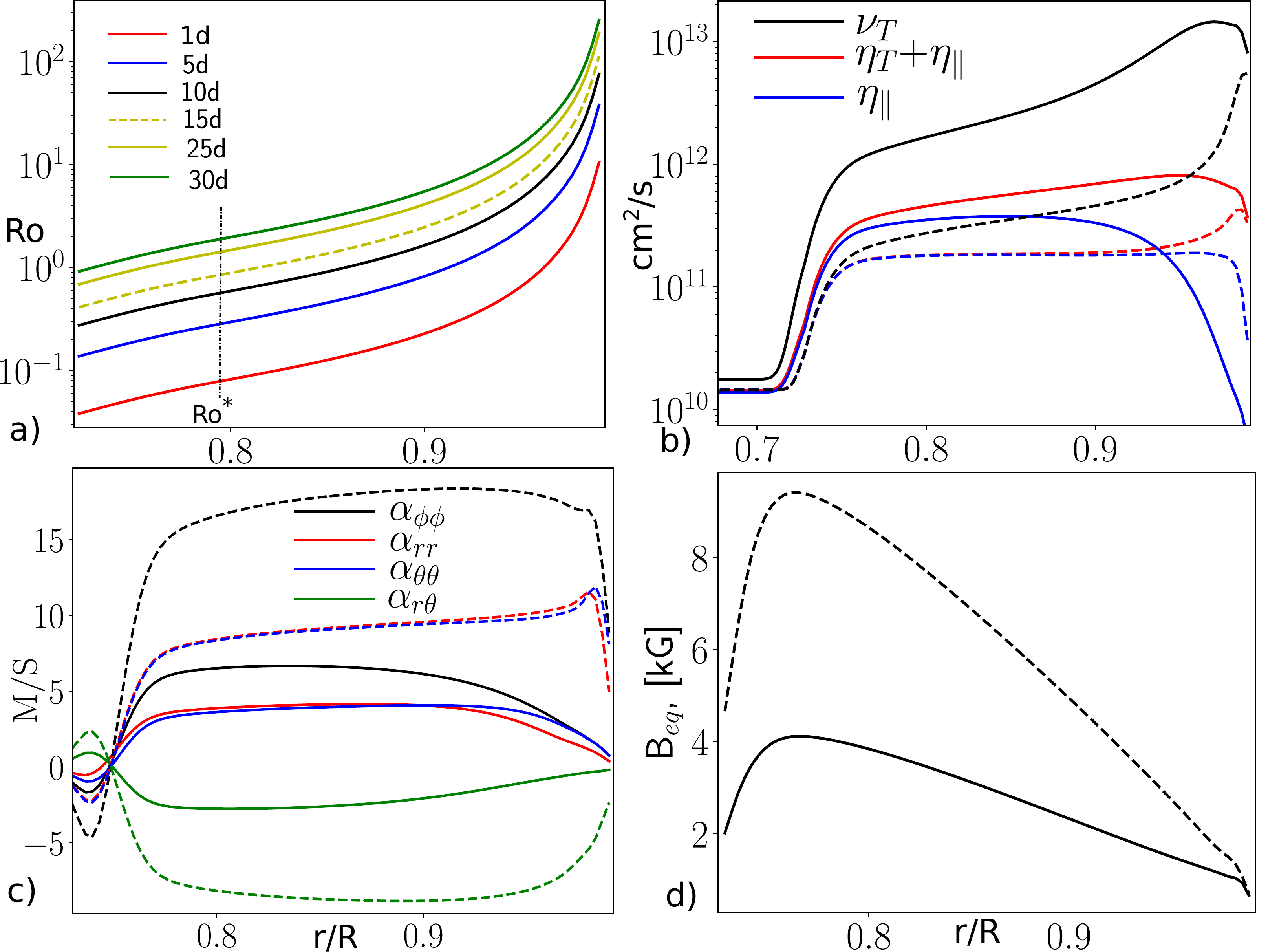}

\caption{\label{rossby}a) The radial profiles of the Rossby number, the line
$\mathrm{Ro}^{*}$ marks the values of the global parameter for each
star (see, the text); the radial profiles for the isotropic eddy viscosity,
$\nu_{T}$ (black line), the total eddy diffusivity, the isotropic
and anisotropic parts, $\eta_{T}+\eta_{\parallel}$, (red line) and
the anisotropic eddy diffusivity induced by rotation, $\eta_{\parallel}$,
(blue line); solid lines for the model M25d and the dashed lines for
the model M2; c) the same as b) for the hydrodynamic part of the $\alpha$-effect
tensor, $C_{\alpha}\alpha_{ij}^{(H)}$, at $45^{\circ}$ latitude;
d) the radial profiles for the equipartition strength of the large-scale
magnetic field, the model M25 - solid line, and the model M2 - dashed
line. }
\end{figure}

\begin{figure*}
\includegraphics[width=0.8\paperwidth]{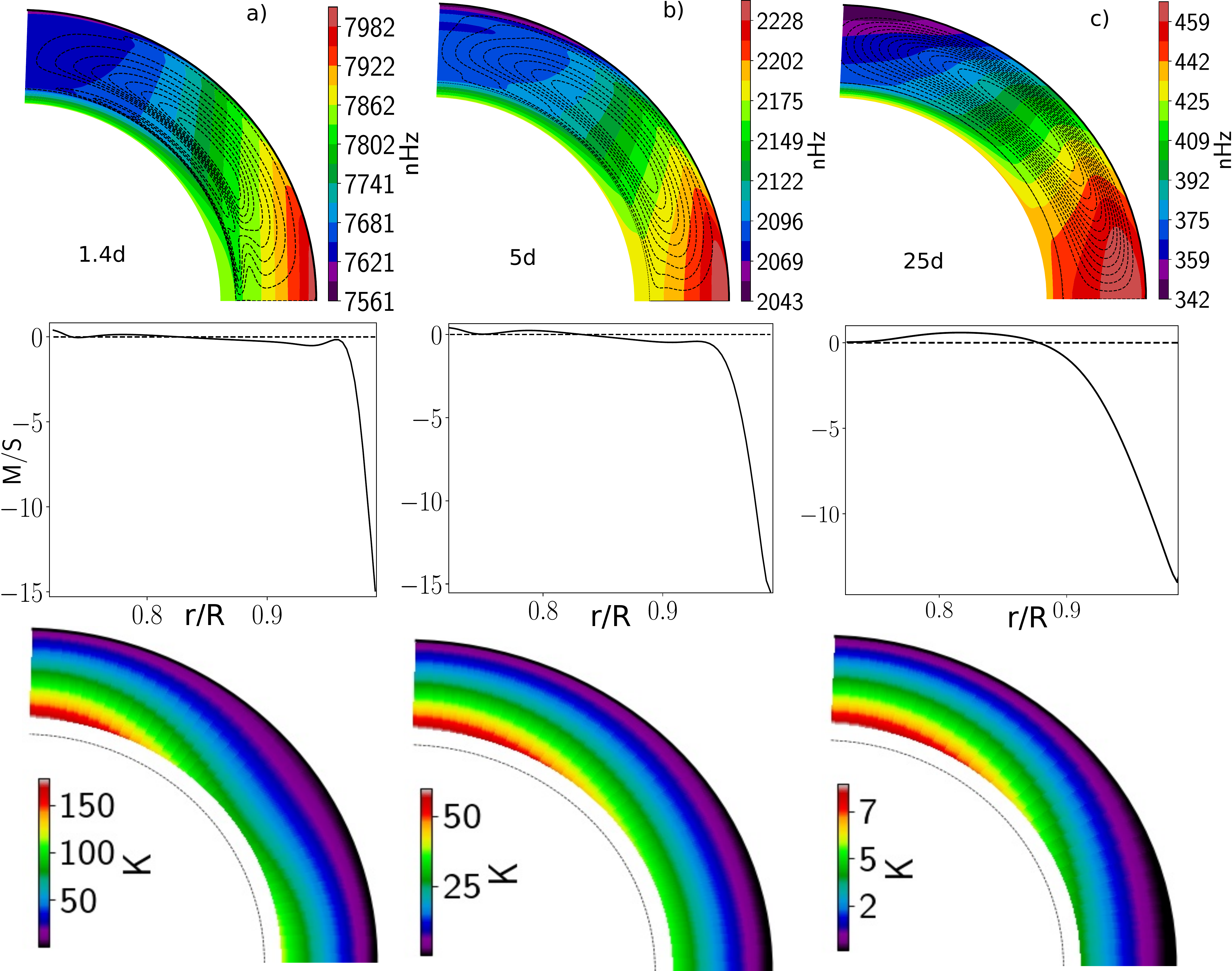}

\caption{\label{fig:dr}The angular velocity (top), meridional circulation
at $45^{\circ}$(middle), and the differential temperature (see, the
Eq\ref{eq:T}) profiles for the star rotating with period: a) 1.4
days; b)5 days c)25 days.}
\end{figure*}

In this paper, we use the same reference convection zone model for
all solar analogs rotating with periods from 1.4 to 30 days.{
}Figure \ref{rossby}a) shows the radial profiles of the Rossby number
for each model. Following \cite{Castro2014} we define the stellar
Rossby number, $\mathrm{Ro}^{*}=\mathrm{P_{rot}}/\tau_{c}$, using
value of $\tau_{c}$ at the distance of one pressure height scale
above the bottom of the convection zone. We find that for the Sun
rotating with a period of 25 days $\mathrm{Ro}^{*}\approx$1.4 which
agrees roughly with the above-cited paper. Figures \ref{rossby} b)
and c) show the radial profiles of the eddy viscosity, eddy diffusivity,
and the kinetic $\alpha$ -effect tensor for the models of the star
rotating with the period of 25 and 2 days. Note, that we use the same
$\alpha$-effect parameter, $C_{\alpha}=0.04$ as in the paper PK19.
We see that effect of rotation results in quenching of the eddy viscosity
and the eddy diffusivity coefficients in the main part of the convection
zone \citep{Kitchatinov1994}. This quenching is, to some extend,
compensated by the effect of rotation on the mean entropy profiles.
The solution of the heat transport equation for the rotating star shows
that the higher the rotation rate, the stepper radial profile of the mean
entropy. In following the mixing length theory, this results to increase
of the convective RMS velocity (see, the Eq\ref{eq:uc}). Therefore
we find that the strength of the equipartition magnetic field in the
model M2 is about factor 2 higher than in the model M25. Note, the
stepper radial profile of the mean entropy in case of the star rotating
with period 2d results in the higher amplitude of the $\alpha$
-effect in comparison with the solar case.

\begin{table*}
\caption{\label{tab}The integral parameters of the kinematic dynamo models.
{The numbers in the run's names correspond to the period of rotation
in days, which is rounded to the nearest integer;} $P_{\mathrm{rot}}$
is the period of stellar rotation; $\mathrm{Ro}^{*}$ - estimation of
the Rossby number, $F_{T}$ is the magnitude of the total magnetic
flux in the convection zone; $F_{S}$ is the magnitude of the magnetic
flux in subsurface layer, $r=0.89-0.99R$; $B_{T}/B_{P}$ is the ratio
between the strength of the toroidal and poloidal magnetic field in
the model; $\beta_{\mathrm{max}}$ is maximum $\mathrm{\beta=\left|\overline{\mathbf{B}}\right|/\sqrt{4\pi\overline{\rho}u_{c}^{2}}}$
and its cycle variations; $P_{\mathrm{cyc}}$ stands for the dynamo
periods found for in the model, the boldface font mark the primary
period of the near-surface low latitude toroidal magnetic field dynamo
waves. The auxiliary periods $P_{PY}$, $P_{D}$ and $P_{\beta}$
are the dynamo periods for the parameters of the Parker-Yoshimura
waves, magnetic diffusivity, and magnetic buoyancy time scales, correspondingly. }

\begin{tabular}{l>{\raggedright}p{0.8cm}>{\raggedright}p{0.8cm}>{\raggedright}p{1.4cm}>{\raggedright}p{1.4cm}>{\raggedright}p{1.9cm}>{\raggedright}p{1.2cm}>{\raggedright}p{1.4cm}>{\raggedright}p{0.8cm}>{\raggedright}p{0.8cm}>{\raggedright}p{0.8cm}}
\hline 
Model  & $\mathrm{P_{rot}}$,

day  & $\mathrm{Ro}^{*}$  & $F_{T}$,{[}MX{]}

$10^{24}$  & $F_{S}$,{[}MX{]}

$10^{24}$  & $B_{T}/B_{P}$  & $\beta_{\mathrm{max}}$  & $\mathrm{P_{cyc}}$,

yr  & $P_{PY}$

yr  & $P_{D}$

yr  & $P_{\beta}$

yr\tabularnewline
\hline 
M1  & 1.38  & 0.08  & 15$\pm$1  & 10$\pm$0.09  & 4248$\pm$443  & 1.3$\pm$0.2  & {1.3}/\textbf{2.7}/7.1  & \textbf{3.3}/4.5  & 51.9  & 0.54\tabularnewline
M2  & 2.5  & 0.12  & 9.5$\pm$0.7  & 6$\pm$0.5  & 1900$\pm$200  & 1.$\pm$0.09  & \textbf{1.5}/2.5/5.2  & \textbf{2.8}/4.2  & 43.1  & 0.65\tabularnewline
M5  & 5  & 0.29  & 4.5$\pm$0.3  & 2.3$\pm$0.2  & 720$\pm$104  & 0.9$\pm$0.1  & 1.9/\textbf{2.5}/15.3  & \textbf{6.3}/6.3  & 32.9  & 0.97\tabularnewline
M8  & 7.9  & 0.29  & 3.$\pm$0.3  & 1.5$\pm$0.2  & 727$\pm$142  & 0.7$\pm$0.07  & 2.4\textbf{/3.1}/9.6  & 6.9  & 30.9  & 2.1\tabularnewline
M10  & 10  & 0.59  & 2.3$\pm$0.1  & 0.8$\pm$0.1  & 596$\pm$42  & 0.4$\pm$0.02  & 2.4\textbf{/3.5}/6.8  & 4.9  & 34.8  & 7.4\tabularnewline
M15  & 15  & 0.88  & 1.5$\pm$0.2  & 0.6$\pm$0.2  & 360$\pm$67  & 0.28$\pm$0.04  & 4.3/\textbf{6.7}  & 7.6  & 29.9  & 15.2\tabularnewline
M20  & 20  & 1.17  & 1.4$\pm$0.1  & 0.45$\pm$0.1  & 384$\pm$106  & 0.24$\pm$0.04  & 8.25  & 11  & 25.3  & 19.6\tabularnewline
M25  & 25  & 1.46  & 1$\pm$0.1  & 0.35$\pm$0.15  & 352$\pm$56  & 0.16$\pm$0.03  & 10.6  & 11.6  & 25.  & 47.1\tabularnewline
M30  & 30  & 1.98  & 0.3$\pm$0.02  & 0.05$\pm$0.02  & 281$\pm$66  & 0.05  & 15.4  & 12.7  & 23.5  & 330\tabularnewline
\end{tabular}
\end{table*}

The angular velocity, meridional circulation, and the relative temperature
profiles for the kinematic versions of the model are illustrated in
Figure \ref{fig:dr}. {The mean over the convection zone Taylor
number,}
\begin{equation}
Ta=\frac{4\Omega_{0}^{2}R^{4}}{\nu_{T}^{2}},\label{eq:Ta}
\end{equation}
{varies in the range from 8.2$10^{6}$ for the period of rotation
of 30 days to 8.3$10^{11}$ for 1.4 days period.} The model of the
differential rotation for the star rotating with a period of 25 days
is the same as reported in PK19. The angular velocity profile in this
model agrees well with helioseismology data. {Note, that the
$\Lambda$-effect in our models is quenched when $r$ is approaching
the bottom of the convection zone. There are two reasons for this.
Firstly, it can be due to the effect of the Coriolis force. This effect
was suggested both in analytical studies (e.g., \citealp{Kitchatinov1993a,Kueker1996,Rogachevskii2018})
and in the global convective simulations of \cite{Warnecke2013b}.
Secondly, the $\Lambda$-effect is saturated because of strong the
radial stratification of the local correlation time of the convective
flows near the bottom of the convection zone (see, \cite{Pipin2018c}).}
{The decrease of the $\Lambda$-effect results in quenching
of the meridional circulation near the bottom of the convection zone.}
A more detailed discussion about mechanisms generating the large-scale
flow can be found in the above-cited paper (also see, \citealp{Kitchatinov1999a,Kitchatinov2005}).
For the star rotating with a period of 1 day, the angular velocity
profile gets close to the cylinder. This is in agreement with the
results of \cite{Kitchatinov2011b}. Note, that the differential rotation
is concentrated on the equator. For the case of the fast rotation,
e.g., the models M1 and M5, the meridional circulation in the depth
of the convection zone is suppressed. {The poleward circulation
is concentrated on the surface. The amplitude of the poleward flow
is increased from 13.5 m/s for the star rotating with a period of 30
days to 15 m/s for the 1.4 days rotating star. }For the star rotating
with a period more than 5 days the rotational profile is close to
the modern Sun, except the magnitude of the differential rotation
is different. {We calculate the differential temperature, 
\begin{equation}
\delta T=T\frac{\overline{s}-\left\langle \overline{s}\right\rangle }{c_{p}},\label{eq:T}
\end{equation}
where $\left\langle \overline{s}\right\rangle $ is the mean density
profile over latitude. The solar model shows the pole-equator difference
of the differential temperature about 7K at the bottom of the convection
zone. It is about 1K at the top of the dynamo domain. This is in qualitative
agreement with the results of \cite{Kitchatinov2011b}. The faster
rotating stars show the higher the pole-equator difference of $\delta T$
. This is due to the rotationally induced anisotropy of the convective
heat transport (\cite{Kitchatinov1999a}). The differential temperature
contrast between pole and equator shows the maximum value at the bottom
of the convection zone. However, the ratio $\delta T/T$ has a maximum
at the top of the convection zone (also, cf, {\citealp{Warnecke2016,Kapyla2019}}).
The contrast of $\delta T/T$ varies from $10^{-6}$ near the bottom
to $10^{-5}$ near the top of the dynamo domain. }

\begin{figure*}
\includegraphics[width=1\textwidth]{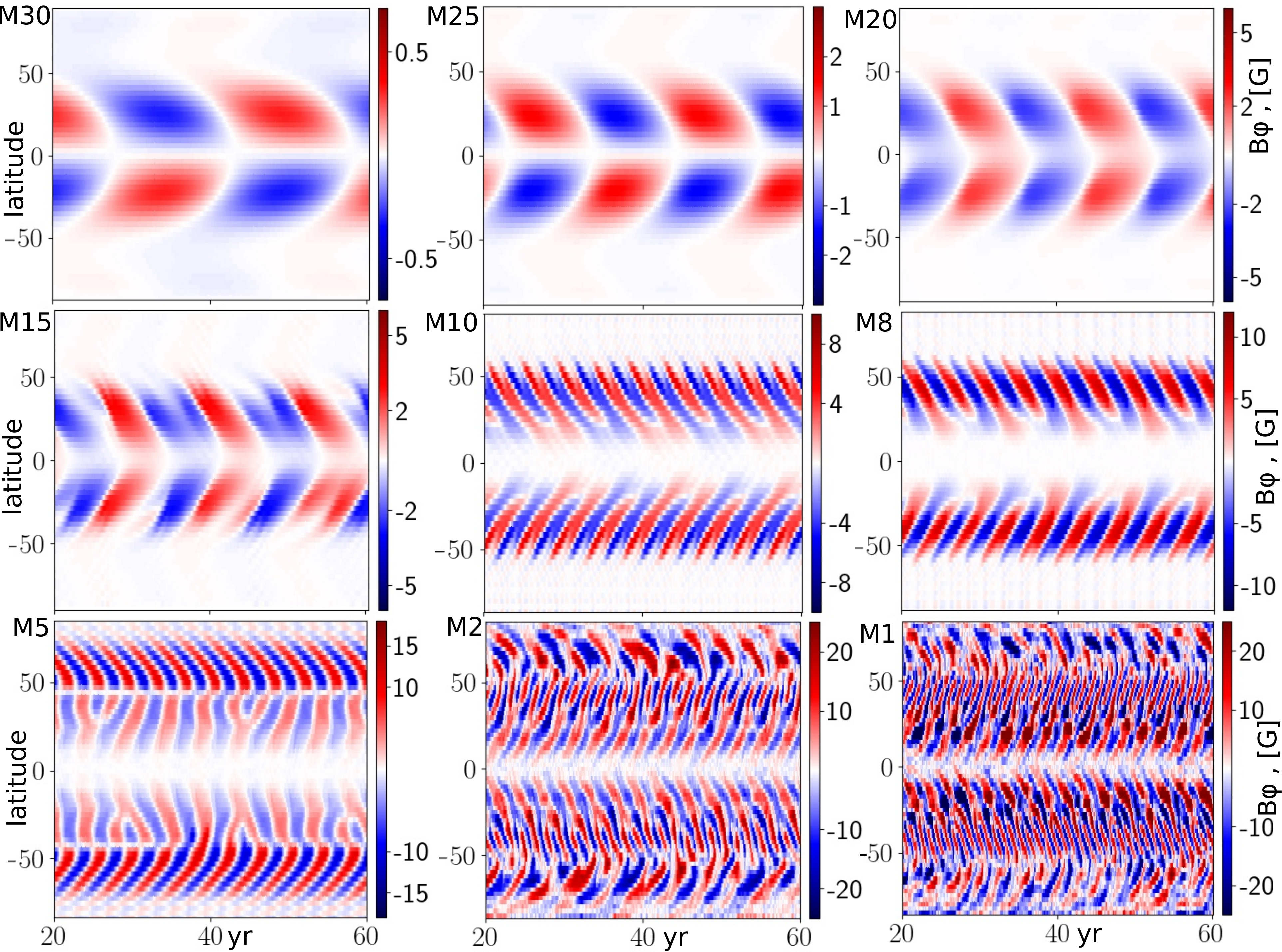}

\caption{\label{fig:kinl}The time-latitude diagrams for the near-surface (r=0.9R)
toroidal magnetic fields for the kinematic dynamo models.}
\end{figure*}

\section{Results}

\subsection{Kinematic models}

As the first step, we discuss the results for the kinematic models
with the nonlinear $\alpha$ and magnetic buoyancy effects.{
In these models, we neglect the magnetic feedback on the large-scale
flow and the mean entropy. In these models, the magnetic field evolution
follows to solutions of the Eq(\ref{eq:mfe-1}) and Eq(\ref{eq:helcon}).
}For the given parameters of the $\alpha$ -effect and the angular
velocity profile the star with the rotation period of 30 days is slightly
above the large-scale dynamo instability threshold. The Table\ref{tab}
lists the integral parameters for the kinematic dynamo models. Our
results are in general agreement with the results of the previous
paper of \cite{Pipin2015}. The models show a decrease in the dynamo
period with an increase in the rotation rate. Figure \ref{fig:kinl}
shows the time-latitude diagrams for the large-scale toroidal magnetic
field evolution at r=0.9R in the kinematic models. The models with
a period of rotation longer than 10 days show the solar-like butterfly
diagrams. {The star rotating with a period of 10 days shows
a double dynamo wave pattern, where the high-frequency high latitude
cycles merge into the long cycles at low latitudes. A similar pattern
is found for the star rotating with the period of 8 days.} {For
star with a higher rotation rate the dynamo wave patterns divide
for two: the high latitudes wave drifts to the equator and the low latitude
waves migrate toward the poles.} This property becomes very clear
in the case of the M5, M2, and M1 models. All those models show a
mix of the magnetic parity modes and different systems of the dynamo
waves at the high and low latitudes. The complicated dynamo wave patterns
in these models result in multiple dynamo periods,
see Table\ref{tab}. For example, {the model M1 shows three
different dynamo waves. There are two high-frequency waves migrating
toward the poles below 50$^{\circ}$. The third one is a quasi-periodic
high-latitude wave migrating toward the equator.} We deduce the dynamo
periods using the time-latitude diagrams of the toroidal magnetic
field in subsurface layer r=0.9R, the variations of the toroidal magnetic
field flux, $F_{S}$, and the Poynting flux luminosity, $L_{P}$.
To identify the dynamo periods we employ the standard wavelet package
of the {\footnotesize{}{}SCIPY }distribution{\footnotesize{}{} (www.scipy.org).}
All those parameters indicate the unique dynamo period for the models
in the range of the rotational periods from 15 to 30 days. The models
for the rotational period of less than 15 start to show the long-term
variations. The different dynamo parameters show the different sets
of periodic variations. For the long period, we choose the minimal
value which is found both in the $F_{S}$ and $L_{P}$ and in the
time-latitude diagrams. Also, the models for the fast rotating star,
M1, M2, and M5, show the short-term periodicity of the $L_{P}$ parameter.
Its period is about twice less of the main dynamo period.

For the fast rotating stars, the range of latitudes with poleward
migration of the dynamo waves corresponds to the extension of the
equatorial super-rotation region. This region shows the cylinder-like
angular velocity profiles. The subsurface shear layer in these models
is shallower than for the slow rotating cases. On the other hand,
the polar regions show the angular velocity profiles which are close
to radial. This results in equator-ward dynamo wave propagation.

\begin{figure}
\includegraphics[width=0.99\columnwidth]{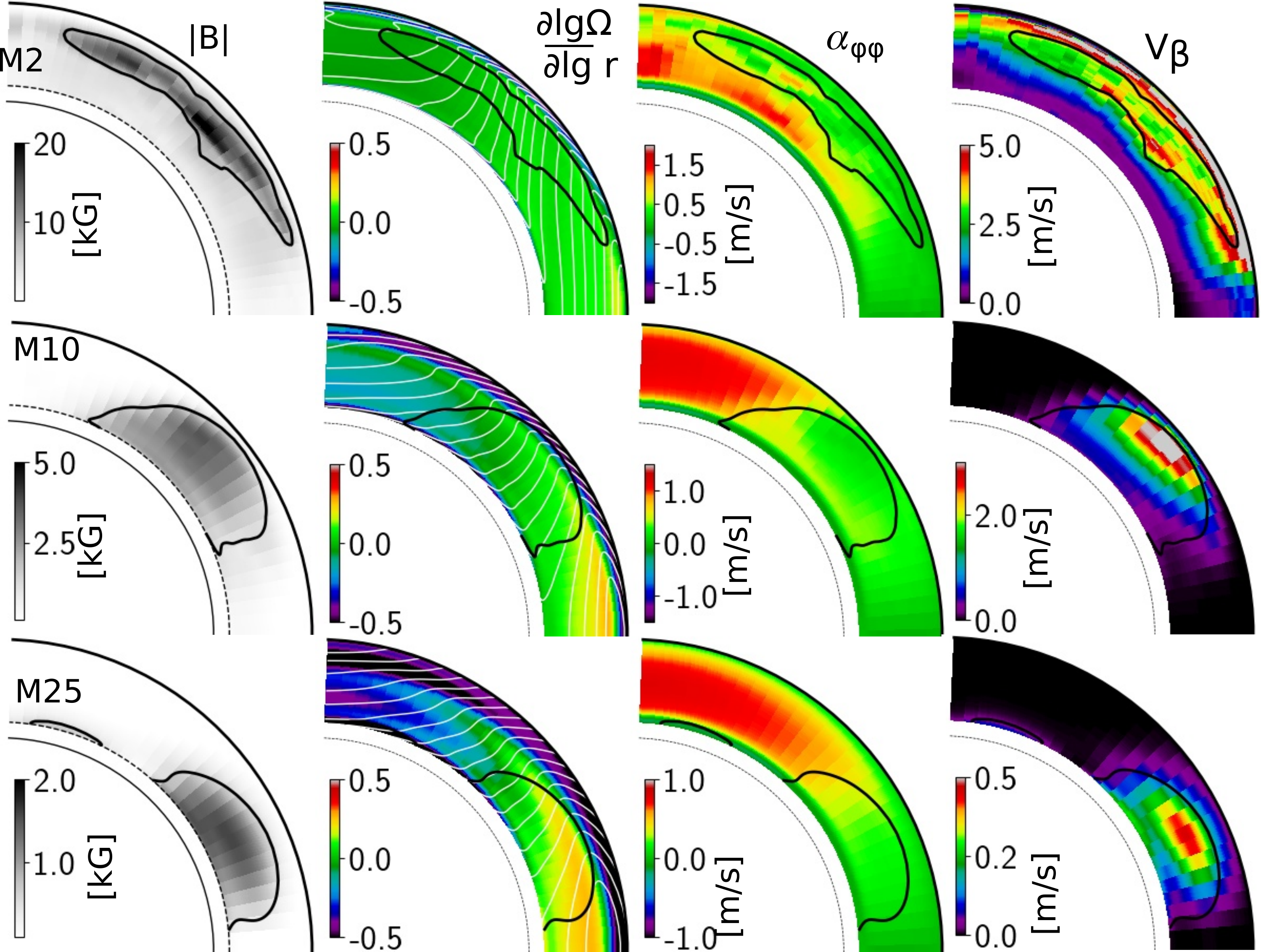}

\caption{\label{kin-sns}{ Snapshots of the models M2, M10, and M25.
The first column shows the run  average  magnetic field strength,
$\left|B\right|$ ; the second column - the mean shear, $\partial\log\Omega/\partial\log r$
and angular velocity profiles; the third column- the mean $\alpha_{\phi\phi}$
(including the magnetic helicity contribution); the fourth column -
the mean magnetic buoyancy velocity. The black line contours show
the position of the $\frac{1}{3}$$\max\left|B\right|$. The points
inside the contour are used to calculate the dynamo periods.}}
\end{figure}

{In general, the dynamo waves patterns show drifts in agreement
with the prediction of the Parker-Yoshimura rule \cite{Yoshimura1975},
i.e., the negative shear and the positive $\alpha$-effect results
in the equator-ward drift.  The Figure \ref{kin-sns} shows snapshots
of the magnetic field strength RMS which are calculated by the time-averaging
over the run. Also, we show the RMS for the total $\alpha_{\phi\phi}$
, the magnetic buoyancy and the mean shear, $\partial\log\Omega/\partial\log r$.
It is found that the increase in the rotation rate results in shifts
of the maximum of the magnetic field strength toward the surface.
The position of the upper boundary of the $\frac{1}{3}$$\max\left|B\right|$
touches the subsurface shear layer, where the gradient of the angular
velocity is negative. The increase of the rotation rate results in
magnetic quenching of the $\alpha$-effect both by means of the algebraic
quenching and magnetic helicity conservation. In all the runs, the
RMS $\alpha$-effect remains positive in the North and negative in
the South hemisphere.  According to the Parker-Yoshimura rule, this
provides the necessary conditions for the equator-ward drift of the
toroidal magnetic field in the upper part of the convection zone.
In the solar case, the additional contributions to the equator-ward
dynamo wave drift are due to the meridional circulation and the turbulent
latitudinal pumping. Results in the Figure \ref{fig:dr}show that
the increase of the rotation rate results in quenching of the
meridional circulation in depths of the convection zone. Similarly,
the magnitudes of the latitudinal pumping and the magnetic diffusivity
are quenched. The quenching of the meridional circulation and the
latitudinal pumping results in shift of the magnetic field strength
RMS distribution toward the high latitudes. }

{The dynamo theory suggests (e.g., \cite{Yoshimura1975,Stix1976,Noyes1984,Parker1984})
that for the kinematic dynamo models the dynamo periods are determined
by the parameters of the eigen dynamo modes (hereafter, $P_{PY}$),
the amplitude of the magnetic diffusivity ($P_{D}$), the amplitude
of the magnetic flux loss, e.g., due to the magnetic buoyancy ($P_{\beta}$)
and amplitude of the meridional circulation (e.g., \cite{Choudhuri1995}).
The latter effect does not contribute much to the dynamo solution because
the meridional circulation at the bottom of the convection zone is
about zero in all our models. According to the approach which was suggested
by \cite{Warnecke2014} (also, see, \citealp{Kapyla2016} and \cite{Warnecke2018A})
we calculate the dynamo periods, $P_{PY}$ and $P_{D}$ using the
parameters of the Parker-Yoshimura waves and the magnetic eddy-diffusivity
parameters. The period $P_{\beta}$ is estimated using the time scale
which follows from the magnetic buoyancy velocity.}

{We proceed as follows. For each run, we compute the RMS of
the large-scale magnetic field strength, the RMS of the $\alpha_{\phi\phi}$
and magnetic buoyancy velocity. To calculate, the parameters $P_{PY}$,
$P_{D}$ and $P_{\beta}$ we choose the mesh points with position
inside the level of the $\frac{1}{3}$$\max\left|B_{RMS}\right|$.
Next, we determine 
\begin{equation}
P_{PY}=\frac{2\pi}{2\omega_{PY}},\label{eq:w}
\end{equation}
where the dynamo frequency, in following \cite{Parker1955}, \cite{Yoshimura1975}
and \cite{Stix1976}, is}

{
\begin{equation}
\omega_{PY}=\left|\frac{\alpha_{\phi\phi}k_{\theta}}{2}r\cos\theta\frac{\partial\Omega}{\partial r}\right|^{1/2},\label{eq:py}
\end{equation}
where $k_{\theta}$ is the latitudinal wave number. In most cases we get one dynamo wave per hemisphere therefore we put $k_{\theta}=\pi/2R$.
Inside the region of the $\frac{1}{3}$$\max\left|B_{RMS}\right|$
we choose the mesh points with the positive $\alpha_{\phi\phi}$ ,
and for them, we compute the $P_{PY}$ separately for two cases. In
one case we use the mesh points where the shear $\partial_{r}\Omega<0$
(for the equator-ward wave propagation) and in another case, we do
the same for the mesh points $\partial_{r}\Omega>0$ (the poleward
wave propagation). The latter case is needed for the models M1, M2
and M5 which show both types of dynamo waves. }

\begin{figure}
\includegraphics[width=0.99\columnwidth]{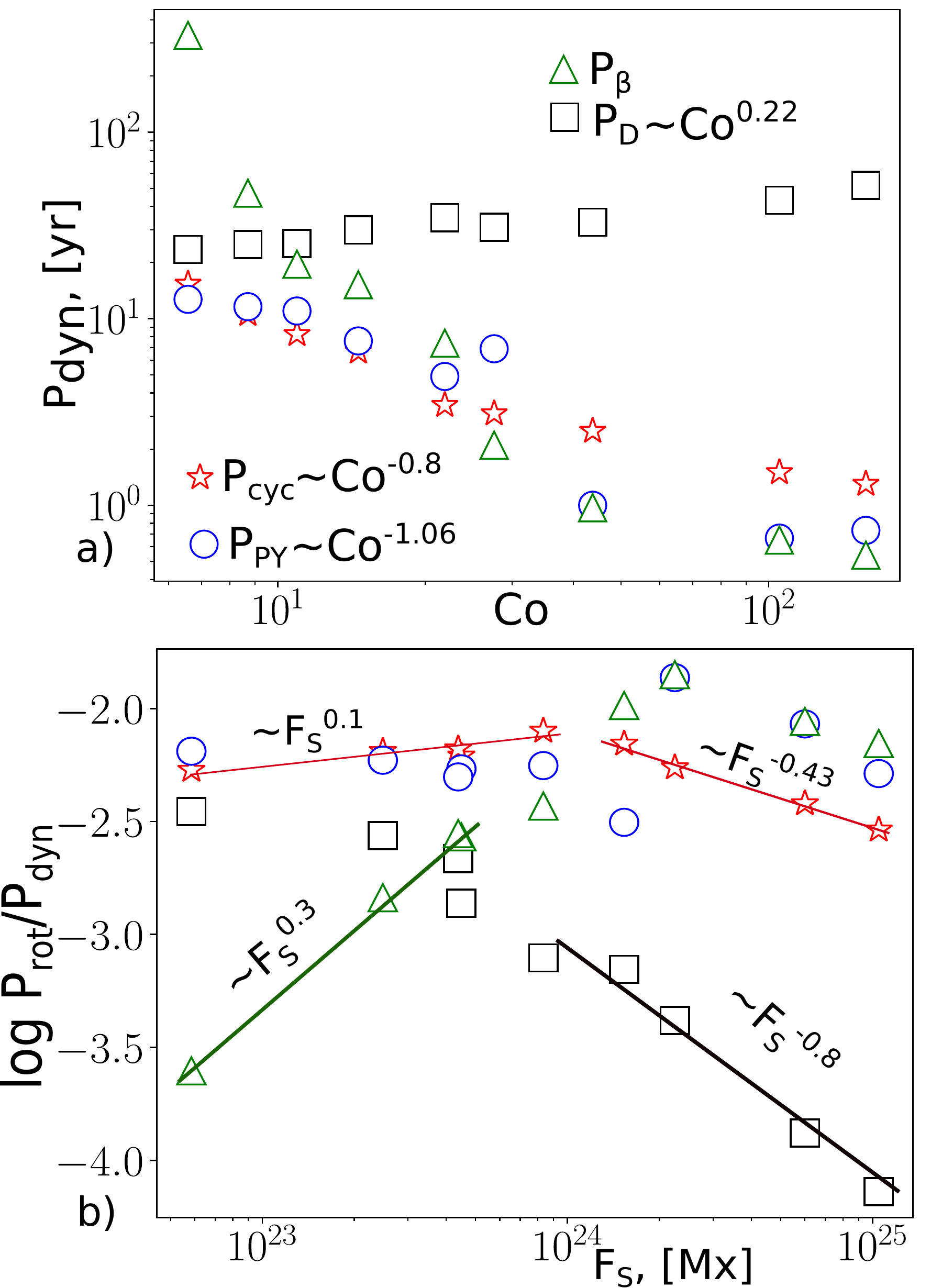}

\caption{\label{fig:perk}a) Relationships of the cycle periods from the Coriolis
number parameter $\mathrm{Co=4\pi/Ro}$; b) the same for the dynamo
frequency normalized to the angular velocity and the magnitude of
the magnetic flux in the subsurface layer. The green and black lines show
the power-law parameters for the magnetic buoyancy and diffusivity
dynamo periods, correspondingly.}
\end{figure}

{The magnetic nonlinearity of the $\alpha$-effect and magnetic
buoyancy results in saturation of the magnetic field growth. In
this regime, the dynamo generation effects are in balance with the
turbulent magnetic diffusion and the magnetic flux loss due to the
magnetic buoyancy. Following to \cite{Roberts1972}, the dynamo period
due to the turbulent diffusion is determined as follows, 
\begin{equation}
P_{D}=\frac{Rd}{\eta_{TS}},\label{eq:eta}
\end{equation}
where $d$ is the thickness of the convection zone and $\eta_{TS}$
is the total magnetic diffusivity at the middle of the convection
zone. For the dynamo period due to the magnetic buoyancy effect we
will use the estimation from dimension arguments, 
\begin{equation}
P_{\beta}=\frac{r_{max}}{\overline{V_{\beta}}},\label{eq:bet}
\end{equation}
where $r_{max}$ is the position of $\max\left|B_{RMS}\right|$ and
$\overline{V_{\beta}}$ is the averaging over the mesh points inside
the level of $\frac{1}{3}$$\max\left|B_{RMS}\right|$. The results
of calculations of the $P_{PY}$, $P_{D}$ and $P_{\beta}$ are shown
in the Table\ref{tab}. The Figure \ref{fig:perk}illustrates the dependencies
of the dynamo cycle periods parameters on the Coriolis number and
magnitude of the magnetic flux in the subsurface layer. It is found that
the dynamo cycle periods in our runs show a good agreement with the
parameter of the Parker-Yoshimura waves. The power-laws for the $P_{cyc}$
and $P_{PY}$ show agreement with calculations of the $P_{PY}$ by \cite{Warnecke2018A} for in the global convection simulations of
the solar-like stars. The time-scale of the magnetic buoyancy loss,
$P_{\beta}$ , shows a rather sharp decrease with the increase of
the rotation rate. Its relationship with the Coriolis number does
not fit into power law. The $P_{D}$ shows the increase of the cycle
period with the increase of the rotation rate. Similarly to \cite{Warnecke2018A}
the relationships of the dynamo period parameters with the magnitude
of the generated magnetic flux reveals the inactive and active branches
of magnetic activity. We return to this analysis after considering
the results of the non-kinematic dynamo models.}

\begin{figure}
\includegraphics[width=0.99\columnwidth]{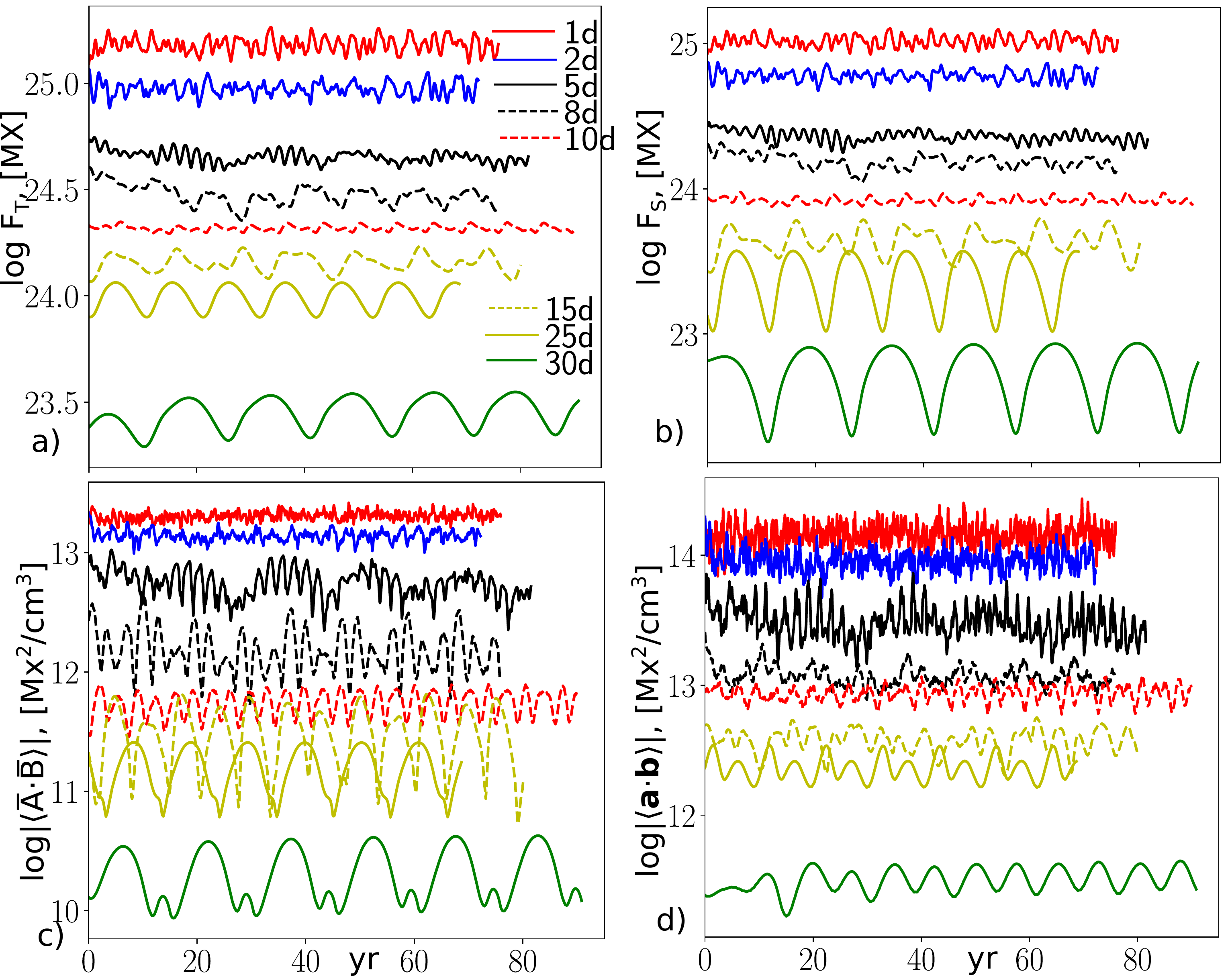}

\caption{\label{kin-en} a) The total toroidal magnetic field flux generated
in the star for the different rotational periods; b) the same as a)
for the total magnetic flux in the subsurface layer $r=0.89-0.99R$;
c) the mean helicity density of the large-scale magnetic field at
the surface; d) the same for the mean small-scale helicity density
at the surface.}
\end{figure}

Figure \ref{kin-en} shows variations of the integral parameters of
the magnetic activity in the kinematic dynamo models. In the kinematic
dynamo models, the decrease of the rotation period from 30 to 1-day
results in an increase of the dynamo-generated toroidal magnetic
field flux by two orders of magnitude. We compute the mean helicity
density of the large- and small-scale magnetic field in the model.
The results are shown in Figures \ref{kin-en} c) and d). We find
that our estimations of the mean helicity density are in agreement
with the results of observations of \cite{Lund2020}. We postpone
their analysis for the next subsections. The small-scale helicity
density shows the order of magnitude larger values than the large-scale
magnetic field helicity density. 
\begin{figure}
\includegraphics[width=0.8\columnwidth]{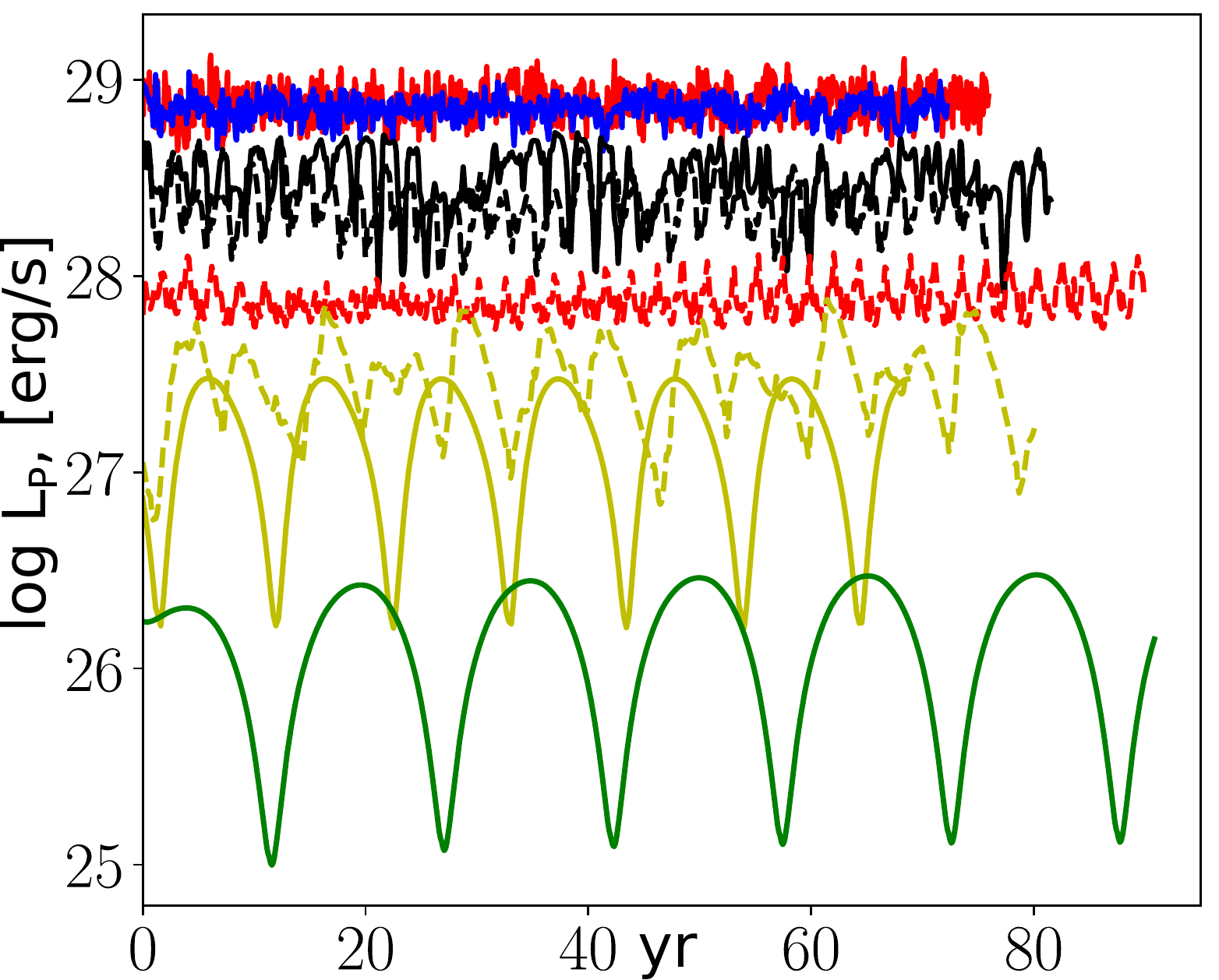}

\caption{\label{kin-poy}The magnitude of the magnetic energy radiated from
the surface. The line notations the same as in Figure \ref{kin-en}.}
\end{figure}

Figure \ref{kin-poy} shows the magnitude of the magnetic energy radiated
from the surface. This parameter shows an increase from $10^{-7}$
of the solar luminosity to $10^{-4}$. The Poynting flux provides
the energy input to the stellar corona. Our values can be used as
an estimation of energy source for the magnetic cycle variation of
the stellar X-ray luminosity. The solar observations show that variations
in the X-ray background flux are the order of $10^{-6}$ \citep{Winter2014}.
Therefore, the estimation of the magnetic luminosity of the model
M25 is enough to explain the solar soft X-ray luminosity variations.
We find that the models M1 and M2 show the saturation for this parameter.

\subsection{Non-kinematic dynamo models}

\begin{table*}
\caption{\label{tab-2}The integral parameters of the non-kinematic dynamo
models. {Here, $\Delta\Omega/\Omega$ is the variation magnitude
of the latitudinal shear at the top boundary, $\pm\mathrm{\delta U_{\phi}}$
is the same for the magnitude of the torsional oscillations,} $\mathrm{\pm\delta U_{\theta}}$
- the same for the magnitude of the meridional flow variations; the
other parameters are the same as in the Table \ref{tab}.}

\begin{tabular}{l>{\raggedright}p{0.7cm}>{\raggedright}p{1.5cm}>{\raggedright}p{0.9cm}>{\raggedright}p{0.9cm}>{\raggedright}p{1.5cm}>{\raggedright}p{1.5cm}>{\raggedright}p{1.5cm}>{\raggedright}p{0.5cm}>{\raggedright}p{1.8cm}>{\raggedright}p{0.7cm}>{\raggedright}p{0.7cm}}
\hline 
Model  & $\mathrm{Ro^{*}}$ & $\Delta\Omega/\Omega$, 10$^{-2}$  & $\pm\mathrm{\delta U_{\phi}}$, m/s  & $\mathrm{\pm\delta U_{\theta}}$, m/s  & $F_{T}$,{[}MX{]}

$10^{24}$  & $F_{S}$,{[}MX{]}

$10^{24}$  & $B_{T}/B_{P}$  & $\beta_{\mathrm{max}}$  & $\mathrm{P_{cyc}}$,

year  & $P_{PY}$

yr  & $P_{\beta}$

yr\tabularnewline
\hline 
M1n  & 0.08  & 0.1$\pm$1.4  & 115.0  & 4.9  & 39.4$\pm$6  & 22.4$\pm$4.  & 1258$\pm$295  & 2.2  & \textbf{3.9}/12.1/21.3  & 0.9  & 0.9\tabularnewline
M2n  & 0.12  & -0.01$\pm$2.8  & 111.3  & 2.9  & 18$\pm$4  & 11$\pm$2.7  & 3793$\pm$869  & 1.9  & \textbf{3.15}/7.8/25.4  & 0.7  & 0.8\tabularnewline
M5n  & 0.29  & 3.6$\pm$1.1  & 127.2  & 11.96  & 4.7$\pm$1.1  & 2.4$\pm$0.8  & 1245$\pm$182  & 1.01  & \textbf{3.4/}18.1  & 2.2  & 2.5\tabularnewline
M8n  & 0.29  & 6.8$\pm$1.4  & 117.5  & 7.1  & 2.2$\pm$0.2  & 1.7$\pm$0.2  & 652$\pm$55  & 0.5  & \textbf{2.1/}4.1/20.9  & 2.6  & 5.2\tabularnewline
M10n  & 0.59  & 12.$\pm$0.3  & 15.2  & 5.1  & 2.1$\pm$0.01  & 0.9$\pm$0.001  & 493$\pm$25  & 0.41  & 2.6  & 3.1  & 6.3\tabularnewline
M15n  & 0.88  & 18.$\pm$0.1  & 8.9  & 2.8  & 1.65$\pm$0.2  & 0.5$\pm$0.15  & 487$\pm$94  & 0.35  & 3.7  & 5.9  & 10.2\tabularnewline
M20n  & 1.17  & 21.2$\pm$0.22  & 5.7  & 1.3  & 1.5$\pm$0.1  & 0.6$\pm$0.1  & 340$\pm$67  & 0.22  & 7.6  & 11.4  & 21.1\tabularnewline
M25n  & 1.46  & 24.8$\pm$0.21  & 3.  & 0.6  & 1.1$\pm$0.1  & 0.37$\pm$0.15  & 301$\pm$75  & 0.21  & 10.1  & 11.9  & 36.9\tabularnewline
M30n  & 1.98  & 31$\pm$0.05  & 0.3  & 0.1  & 0.3$\pm$0.01  & 0.1$\pm$0.01  & 238$\pm$90  & 0.09  & 12.8  & 13.2  & 145\tabularnewline
\end{tabular}
\end{table*}

\begin{figure*}
\includegraphics[width=1\textwidth]{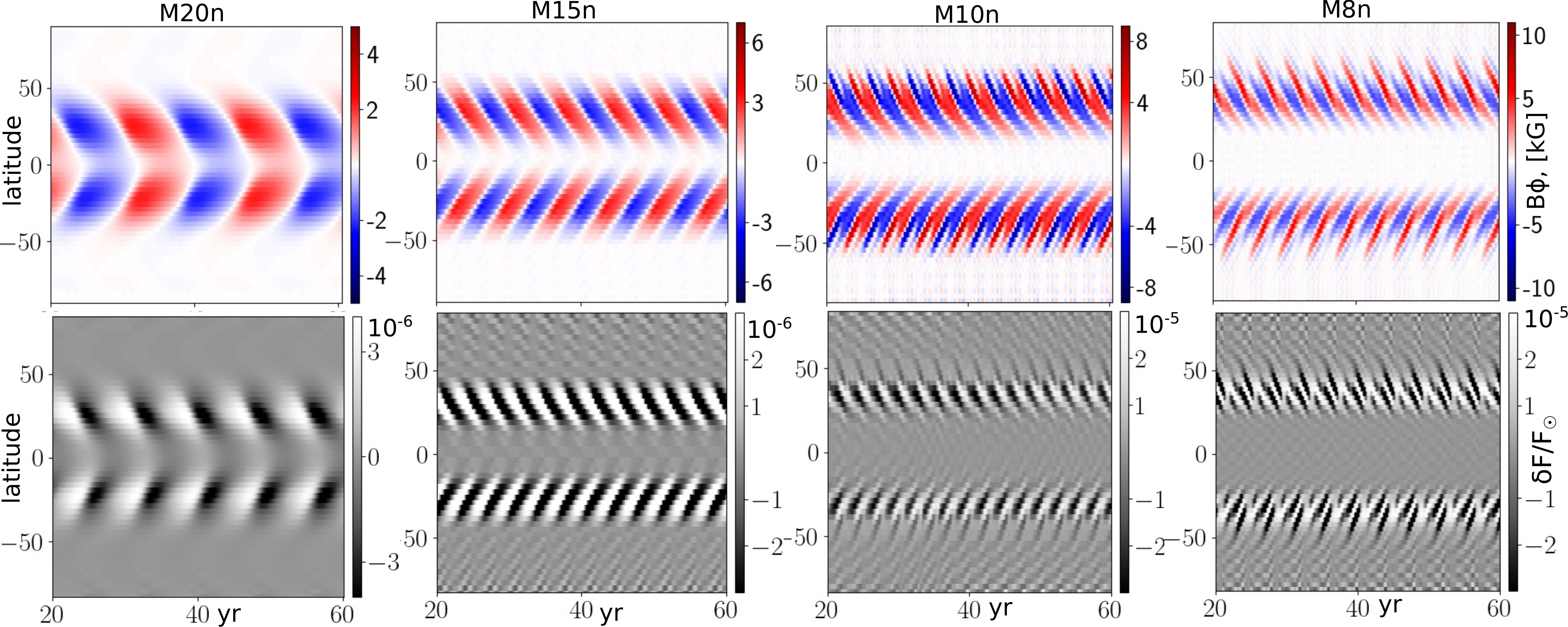}

\caption{\label{fig:nkinl-1}The top row shows the time-latitude diagrams for
the near-surface (r=0.9R) toroidal magnetic fields for the non-kinematic
dynamo models M20n, M15n, M10n, and M8n. The bottom show the same for
the relative variation of the radiation flux $\frac{\delta F_{r}}{F_{\odot}}$,
see the Eq(\ref{eq:dfr}).}
\end{figure*}

{In the non-kinematic dynamo models, we take into account the
magnetic feedback on the large-scale flow and the mean entropy. }Table
\ref{tab-2} lists the integral parameters of the non-kinematic dynamo
models. The non-kinematic models M25dn and M20dn hold the qualitative
properties of the magnetic field evolution the same as their kinematic
versions. The patterns of the torsional oscillations and meridional
circulation variations are qualitatively similar to the results of
\cite{Pipin2019c}. The maximum magnetic field strength in these models
is below half of the equipartition value.  Figure \ref{fig:nkinl-1}
shows the time-latitude diagrams of the near-surface toroidal magnetic
field in the non-kinematic runs for the rotational periods' interval
between 8 and 20 days. {In all those runs the toroidal magnetic
field drifts to the equator. The runs for the rotational period shorter
than 25 days show the mixed parity solution. In particular, the quadrupole
parity dominates in the runs M20n and M15n. In the run M10n, the dipole
parity solution dominates and the run M8n shows the quadrupole solution.
The dynamo periods in these runs are shorter than in the kinematic
case. The same is found for the models M30n and M25n. Variations of
the magnetic activity and the large-scale flow result in variations
of the surface radiation flux. The diagrams show the relative variations,
$\delta F_{r}/F_{\odot}$, see the Eq(\ref{eq:dfr}). Results of the
runs M30n and M25n are qualitatively similar to \cite{Pipin2004}
and \cite{Pipin2018b}. Therefore, we do not illustrate it here. Also, the
magnitude of the parameter $\delta F_{r}/F_{\odot}$ in the M25n is
the order of $10^{-6}$, and it is $10^{-8}$ in the run M30n. It is found
that the radiation flux is suppressed during the maximum of the magnetic
cycle. This is due to the magnetic quenching of the convective heat
flux. The period after a maximum of the magnetic cycle shows the relative
increase of the radiation flux. Despite the high amplitude of the
near-surface toroidal magnetic field in the models M8n and M10n, which
is the order of 8kG in the run M8n, the variations of the parameter $\delta F_{r}/F_{\odot}$
in the runs are less than $10^{-4}$. It is much less than that found
in the solar observations. In the model M20n (as well as in M25n),
we find that duration of the relatively high radiation flux is longer
than the duration of the suppressed period. }

{The results for the model M15n are drastically different from
the kinematic case. The non-kinematic model shows the equator-ward
propagating dynamo waves. Their frequency is about as twice as high
in comparison to the model M15. We investigated the origin of the
change of the magnetic butterfly diagrams in the model M15n in some
details. We made additional runs where we alternately switched off
the magnetic effects on the mean heat transport. This results in
a solution that is very similar to the kinematic case. Previously,
in the paper by \cite{Pipin2019c}, we find that the magnetic quenching
on the convective heat transport makes a profound effect on the meridional
circulation. Therefore, the change in the meridional circulation structure,
which is induced by the dynamo, can affect the dynamo evolution,
in particular for the high activity regime.}

\begin{figure*}
\includegraphics[width=1\textwidth]{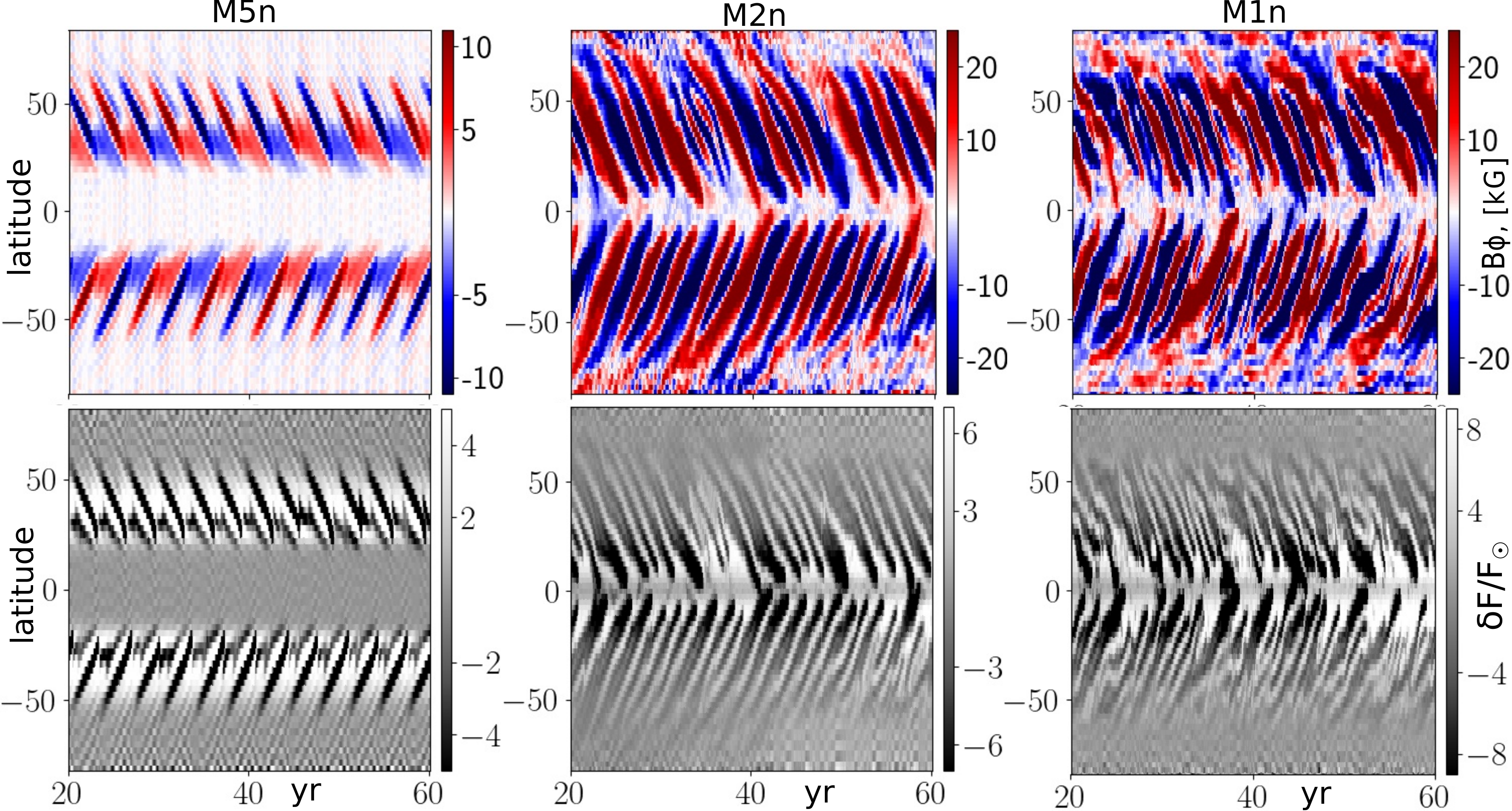}

\caption{\label{fig:nkinl}The same as Figure \ref{fig:nkinl-1} for models
models M5n, M2n and M1n.}
\end{figure*}

The equatorward propagating dynamo waves are found in models M1n,
M2n, and M5n, as well. However, the dynamo period in these models
is longer than in the kinematic case. Figure \ref{fig:nkinl} shows
results for these models. The model M5n shows a qualitatively similar
evolution as the model M8n. The models M1n and M2n show longer
dynamo periods than the models M5n and M8n. These models show the
long-term suppression of the radiation flux because periods of high
magnetic activity dominate. 

\begin{figure*}
\includegraphics[width=0.9\textwidth]{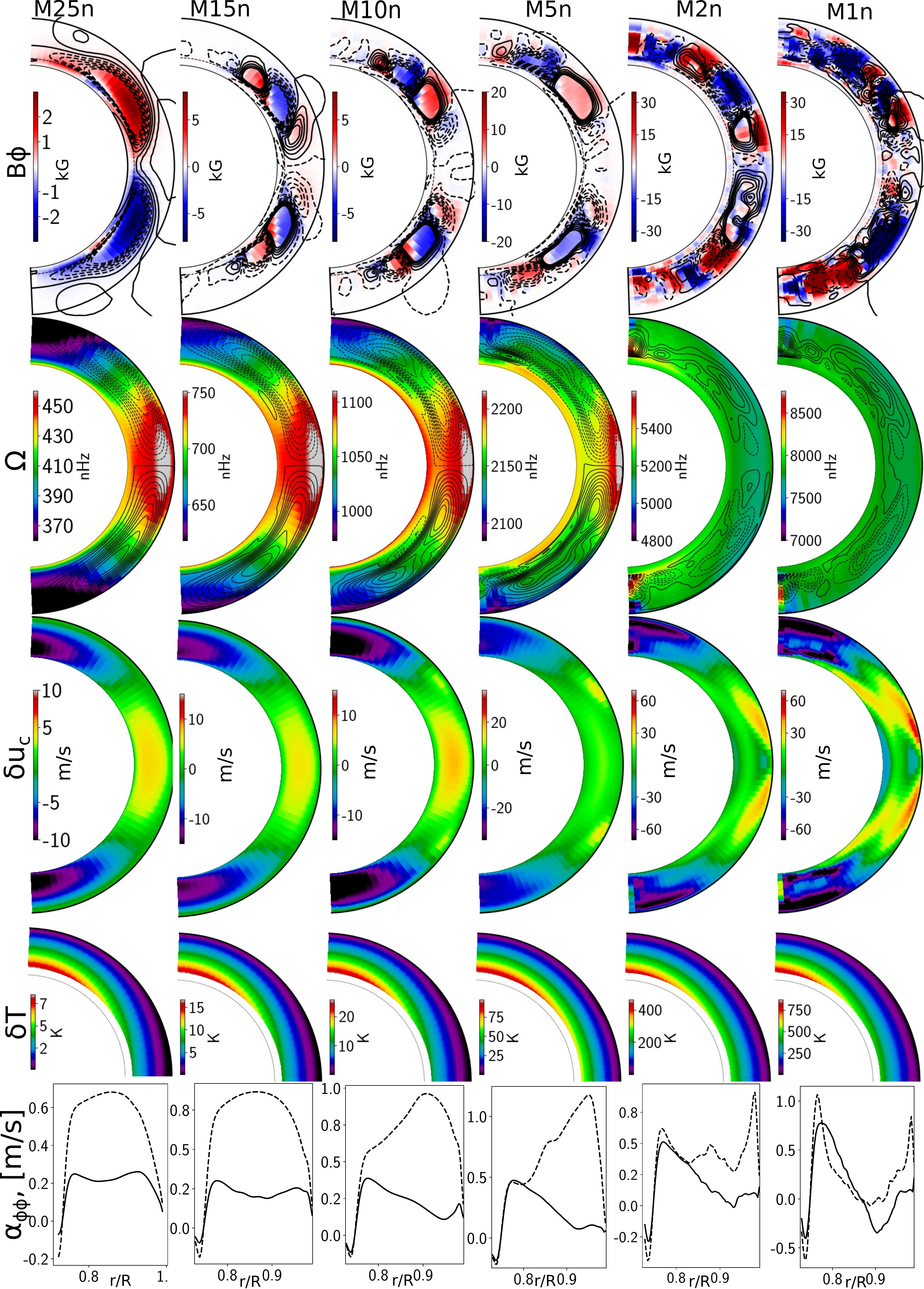} \caption{\label{fig:cvrms}The first row shows snapshots of the toroidal magnetic
field distribution (color image) and streamlines of the poloidal magnetic
field for the non-kinematic dynamo models; the second row shows the
mean over one magnetic cycle the angular velocity distributions
and streamlines of the meridional circulation; the third row shows
the same for the convective RMS perturbation, where the reference
RMS is calculated by averaging over latitudes; the bottom row shows
the same for the mean $\alpha$-effect profiles at the latitudes 30$^{\circ}$
and 60$^{\circ}$. }
\end{figure*}

For the range of the rotational periods from 1 to 10 days, we find
that the major effect of the dynamo on the large-scale flow is the
multi-cell meridional circulation. The models M10n and M15n are at
the boundary of nonlinear bifurcation of one meridional cell into
a multi meridional cell convection zone. Figure \ref{fig:cvrms} shows
the typical snapshots of the magnetic field distributions, the mean
large-scale flows, the differential temperature, and the $\alpha$
effect profiles in the set of the non-kinematic dynamo models. In
our previous paper, we found that the magnetic effects on heat transport
can produce variations of the azimuthal flow and meridional circulation.
When the magnetic field strength is much less the equipartition value,
the magnetic effect on convection is not strong. In this case, variations
of the large-scale flow about the reference state are relatively small.
For the case of weak magnetic activity, the reference profiles of
the large-scale flow are determined by the effect of the Coriolis
force on convection. Effect of rotation on the convective heat transport
results in the difference between the temperature at the stellar equator
and pole. For qualitative analysis, it is useful to see the magnitude
of the convective velocity RMS variations due to the large-scale dynamo.
We compute the relative deviations of the convective RMS velocity
for each model. These deviations are calculated from the mixing-length
expression Eq(\ref{eq:uc}). For the reference state, we use the averaged
over latitude profile of the mean entropy. The average is done for
each snapshot. The results are illustrated in the third row of Figure
\ref{fig:cvrms}.

In the model M25n, the magnetic perturbations of convection flux are
small. In this case, the relative deviations of the convective velocity
RMS is determined by the effect of the Coriolis force. Convection
in the polar regions is suppressed relative to the equator. The effect
is strong near the bottom of the convection zone and it is quenched
toward the surface. So the temperature inhomogeneity is rather small
at the solar surface \citep{Beckers1960,Miesch2008,Teplitskaya2015}.
We find that the models for the rotation periods from 1.4
to 10 days show the stronger latitudinal contrast for the convective
velocity RMS variations than the model M25n. These stars have a higher
rotation rate and a stronger effect of the Coriolis force than the
solar case. We find that the increase of the magnetic activity results
in an increase of the inhomogeneity of the convective velocity RMS
variations, especially, in the near-equatorial region. {This
is seen in the fourth row of the Figure \ref{fig:cvrms}. Interesting
that the temperature contrast between the pole and equator in the
non-kinematic models for the rotational periods from 1.4 to 5 days
is higher than for the kinematic ones. For example, the temperature
contrast at the bottom of the convection zone in the model M1n is
about factor 5 higher than that in the model M1. Both the mean-field
models (\citealp{Kitchatinov1995,Kitchatinov1999a}) and the global
convection simulations (\citealp{Miesch2006,Warnecke2016,Kap2019}
) show that the temperature contrast affects the magnitude of the
differential rotation and meridional circulation structure.} In the
model M15n, there is a weak meridional cell of opposite sign at the
bottom of the convection zone. Besides, the main anticlockwise (in
the North) cell is divided into two cells. It has two center-type stationary
points and one saddle-type stationary point. The same is found for
the models M8n and M5n. In these models, the location of the saddle-type
stationary point corresponds to the local maximum of the convective
velocity RMS variations. In the models M2n and M1n, the bottom clockwise
cell (in the North) becomes dominant. These runs show a strong depression
of the differential rotation in the main part of the convection zone.

The non-kinematic models M1n, M2n, and M5n show the high magnetic
activity with the mean strength of the large-scale magnetic field
exceeding the equipartition value, $\beta\ge$1. There are about 4
activity nests of the toroidal magnetic field in each hemisphere.
The strong magnetic field results in a high deviation of the mean
entropy distribution from the pure hydrodynamic state. From Figure
\ref{fig:cvrms} we find that the maxims of the convective velocity
RMS variations are located in the upper part of the convection zone.
The differential rotation in these models is weak. The averaged angular
velocity profile shows an accelerated rotation regions in the polar
caps. This seems for the first time to be found in the mean-field
dynamo models (cf, \citealp{Kitchatinov1999a}). In the model M2n,
the mean circulation structure in the main part of the convection
zone is opposite to the solar case, i.e., the main circulation cell
is clockwise. Near the surface, there is a weak anti-clockwise circulation
cell.

The mean $\alpha$-effect is, in general, positive for all models.
The models M1n and M2n show inversion of the $\alpha$-effect in the
near-equatorial regions. This inversion is due to the magnetic helicity
conservation \citep{Pipin2013c} and magnetic saturation of the $\alpha$
-effect due to effects of the large-scale magnetic field on the convective
motions. The distributions of the $\alpha$-effect, the angular velocity
profiles together with the meridional circulation provide the equatorward
propagation of the dynamo waves in the upper part of the convection
zone.

{We calculate the parameters of the Parker-Yoshimura dynamo
waves for the non-kinematic dynamo models, as well. The results are
given in the Table \ref{tab-2}. For the case of the star rotating
with a period less than 10 days the period parameters, $P_{PY}$ are
very different from the period, $P_{cyc}$, which we find from the
numerical solution. According to the results presented in Figure \ref{fig:cvrms}
and \ref{2d}, the equatorward drift of the toroidal magnetic field
in the upper part of the convection zone is supported both by the
Parker-Yoshimura rule and the effective drift which is produced by
the meridional circulation and turbulent pumping. Indeed, the models
M1n, M2n, and M5n show the negative gradient of the angular velocity
in the bulk of the convection zone for the latitudes higher than 30$^{\circ}$.
The $\alpha$ -effect shows a positive sign above $0.9R$ in all
those models. Also, we find that the effective equatorial drift velocity
of the large-scale magnetic field is about 5 m/s at the level of $0.9R$.
The butterfly diagrams (see, Figure \ref{fig:nkinl}) show the spatial
length of the toroidal magnetic field dynamo wave of about $\pi R/4$.
The magnitude of the effective velocity drift suggests that the dynamo
wave propagation period is about 3.4 years. This approximately corresponds
to the $P_{cyc}$, which we find in the models M1n, M2n, and M5n. } 
\begin{figure}
\includegraphics[width=0.95\columnwidth]{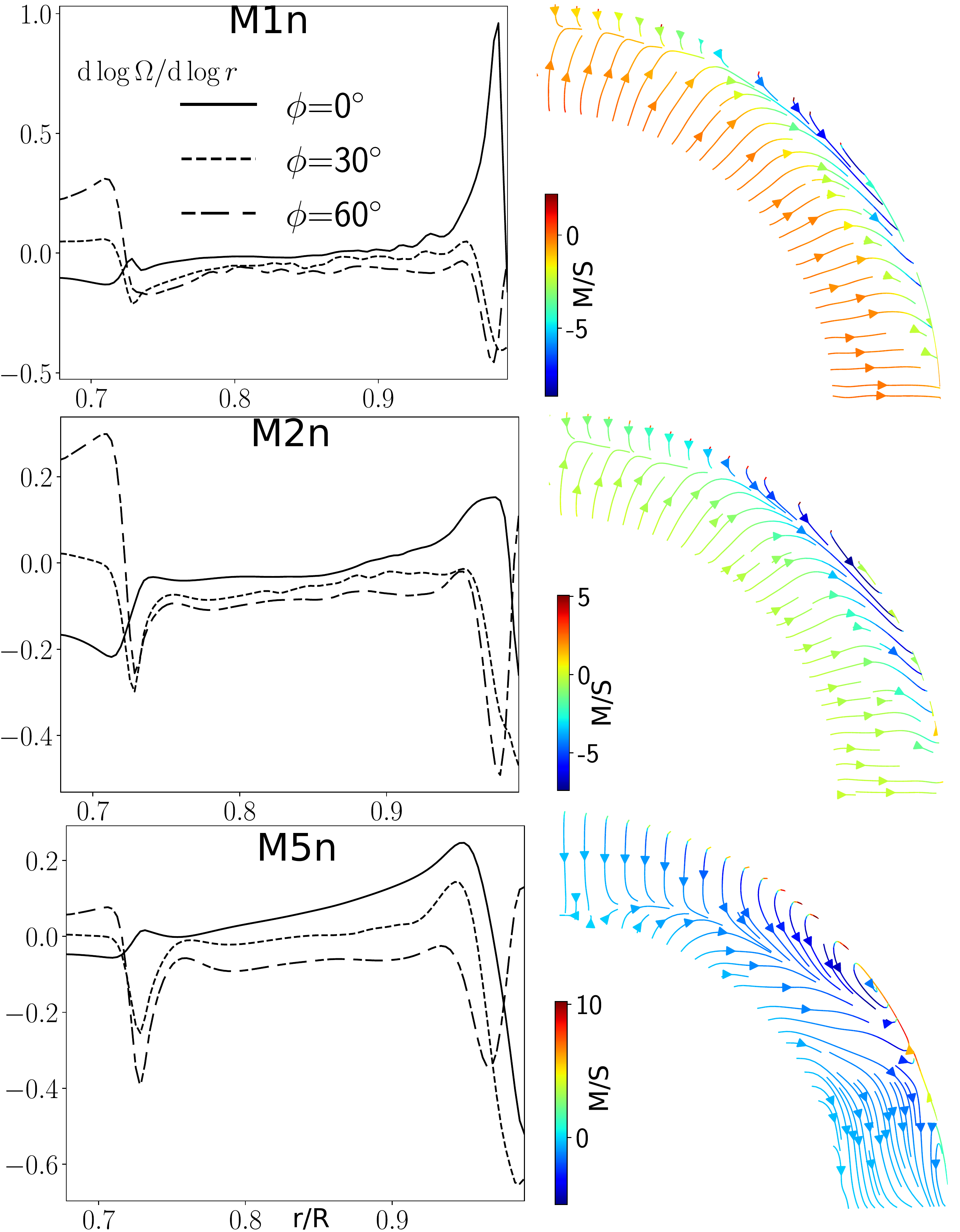}

\caption{\label{2dcf}The left column shows the mean radial profiles of the
angular velocity radial gradient in the non-kinematic dynamo models
M1n, M2n, and M5n at the latitudes $0^{\circ}$, $30^{\circ}$ and
$60^{\circ}$. The right column show streamlines of the mean effective
drift of the toroidal magnetic field, the velocity drift includes
the turbulent pumping due to the density gradient, the magnetic buoyancy, and the meridional circulation velocity. Variations of the color show
the latitudinal effective drift velocity.}
\end{figure}

\begin{figure}
\includegraphics[width=0.95\columnwidth]{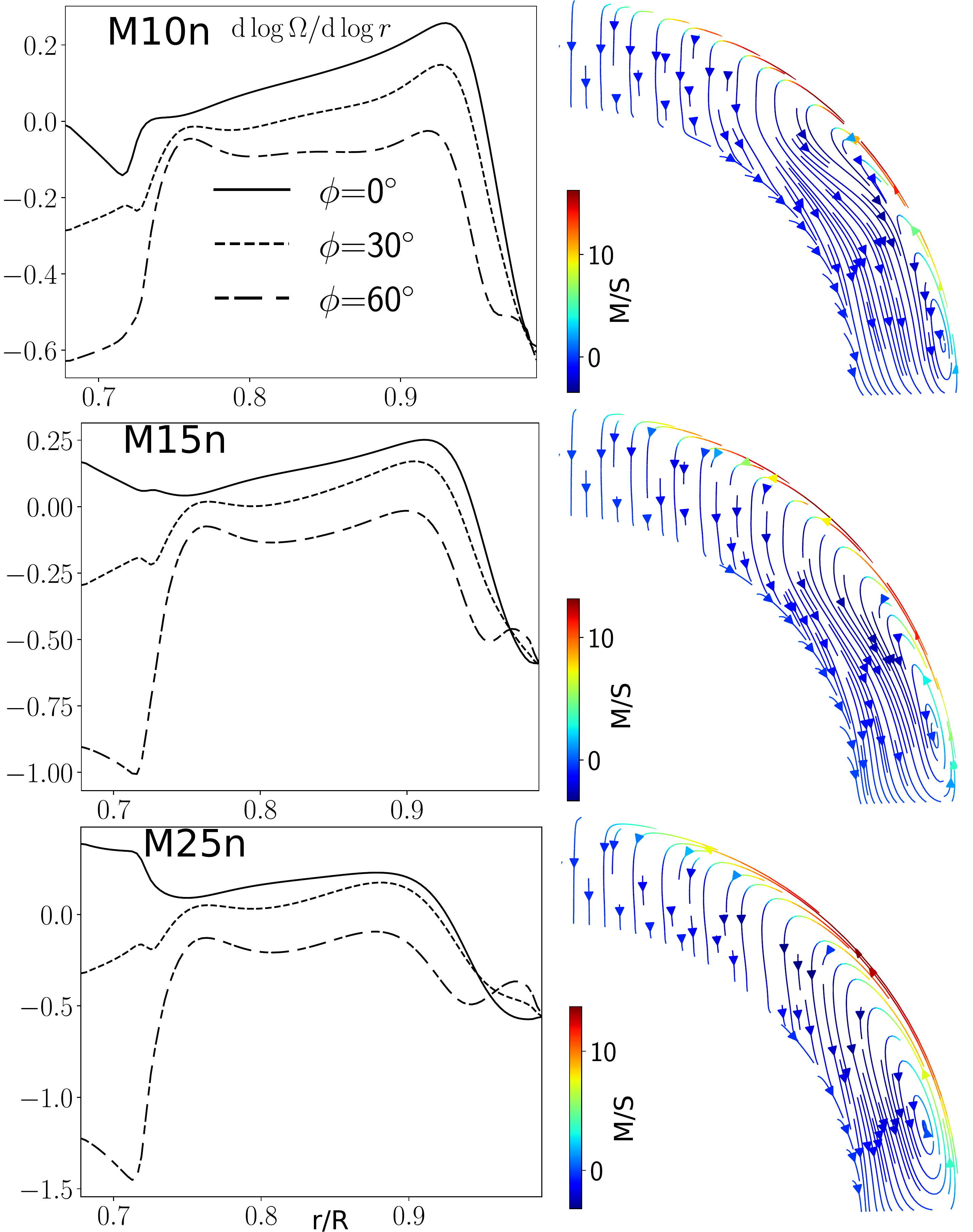}

\caption{\label{2dcf-1}The same as Figure \ref{2dcf} for the models M10n,
M15n and M25n.}
\end{figure}

{Figures \ref{2dcf} and \ref{2dcf-1} show that the decrease
of the rotation rate and magnetic activity results in redistribution
of the effective velocity drift of the toroidal magnetic field in
the bulk of the convection zone. For the fast rotating case, the magnetic
buoyancy becomes dominant. In those cases, the turbulent latitudinal
pumping, as well as the meridional circulation, are quenched in the
lower part of the convection zone. The rotational quenching of the
latitudinal pumping in the upper part of the convection zone is only
moderate because the Rossby number there is higher than near the bottom
of the convection zone. Therefore the models M1n, M2n, and M5n show
the equator-ward propagation in the upper part of the convection zone.
In the opposite case of the slow rotation rate, e.g., the model M25n,
the equator-ward propagation waves at the level of $0.9R$ are supported
both by the Parker-Yoshimura rule for the layer above $0.9R$ and
the effective velocity drift for the layer below $0.9R$.}

\begin{figure}
\includegraphics[width=0.99\columnwidth]{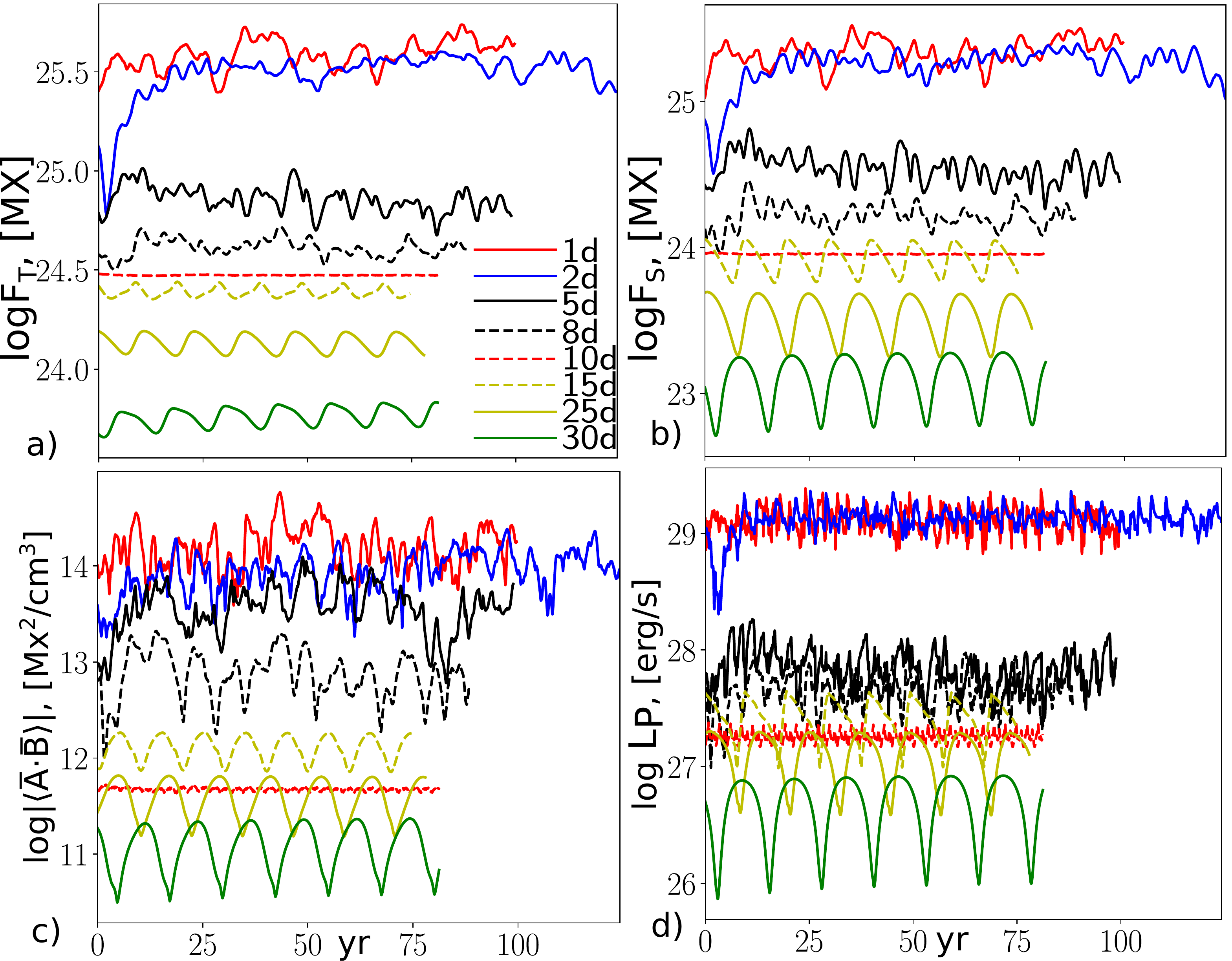}

\caption{\label{intnk}The same as Figure \ref{kin-en} for the non-kinematic
dynamo models.}
\end{figure}

Figure \ref{intnk} shows variations of the integral parameters for
the non-kinematic dynamo models. It is found that the magnitude of
the magnetic activity in the models for the star's rotation period
in the range of 10 to 25 days is decreased in comparison with the
kinematic case. This result is in agreement with our previous calculations
\citep{Pipin2015,Pipin2016b}. Surprisingly, the non-kinematic models
for the stars rotating with periods of 1, 2, and 5 days show a higher
activity level than their kinematic analogs. This is due to a considerable
reorganization of the large-scale flow in these models. Moreover the
models M1dn and M2dn have a week anti-solar differential rotation
in the depth part of the convection zone. The kinematic theory predicts
the anti-solar differential rotation for the slow rotating stars,
which have a high Rossby number \citep{Kitchatinov1995,Kapyla2011,Guerrero2013,Gastine2014,Brandenburg2018,Ruediger2019}.
On the other hand, \cite{Kitchatinov2004b} showed that anti-solar
differential rotation can be generated by means of the magnetically
induced anisotropy of the heat-transport inside the convection zone.
Variations of the magnetic activity parameters in the models for the
rotational periods 8 and fewer days are non-stationary. We find that
in these models the magnetic cycle is well seen in the butterfly diagrams
of the toroidal magnetic field and in the integral parameters of the
magnetic activity including the Poynting flux luminosity and the radiation
flux variations. In variations of the total magnetic flux, the short
cycles can disappear from time to time. Besides, this situation happens
in the time series of the unsigned magnetic flux in the subsurface
layer $r=0.89-0.99R$. Such periods are characterized by the change
of the magnetic parity. The strong parity variations can cause the
variability of the primary cycle length as well.

\begin{figure}
\includegraphics[width=0.5\textwidth]{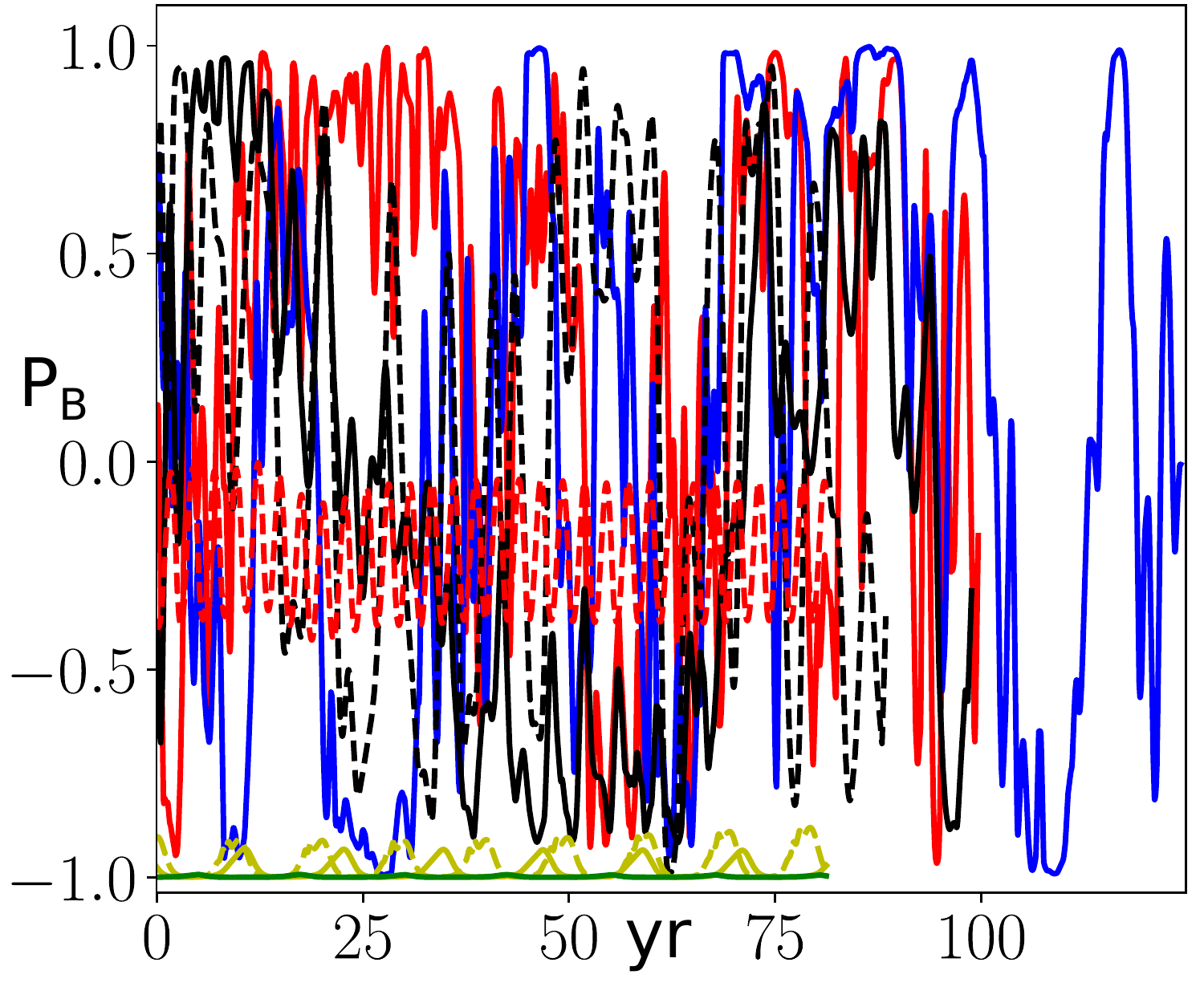}

\caption{\label{par}The magnetic parity parameter in the non-kinematic dynamo
models. The line notation is the same as in Figure \ref{intnk}.}
\end{figure}

Figure \ref{par} shows the evolution of the parity symmetry about
the solar equator in the non-kinematic dynamo models. The models for
the rotational periods from 15 to 30 days show $P_{B}\approx-1$,
which means that the antisymmetric about the equator toroidal magnetic
field dominates. The models for the range of periods from 1 to 8 days
show the long-term variations of the $P_{B}$ from dipole to quadrupole
type symmetry and back. The model M10n shows $P_{B}\approx0$. We
see that despite time-latitude variations of the toroidal magnetic
field, this model shows no variations of the integral parameters $F_{T}$,
$F_{S}$ and it shows a small magnitude variation of the mean large-scale
magnetic helicity density and the Poynting flux luminosity. The mean
values of the integral parameters are slightly less than in the kinematic
model M10.

\subsection{Rotation-activity relations}

\begin{figure}
\includegraphics[width=0.5\textwidth]{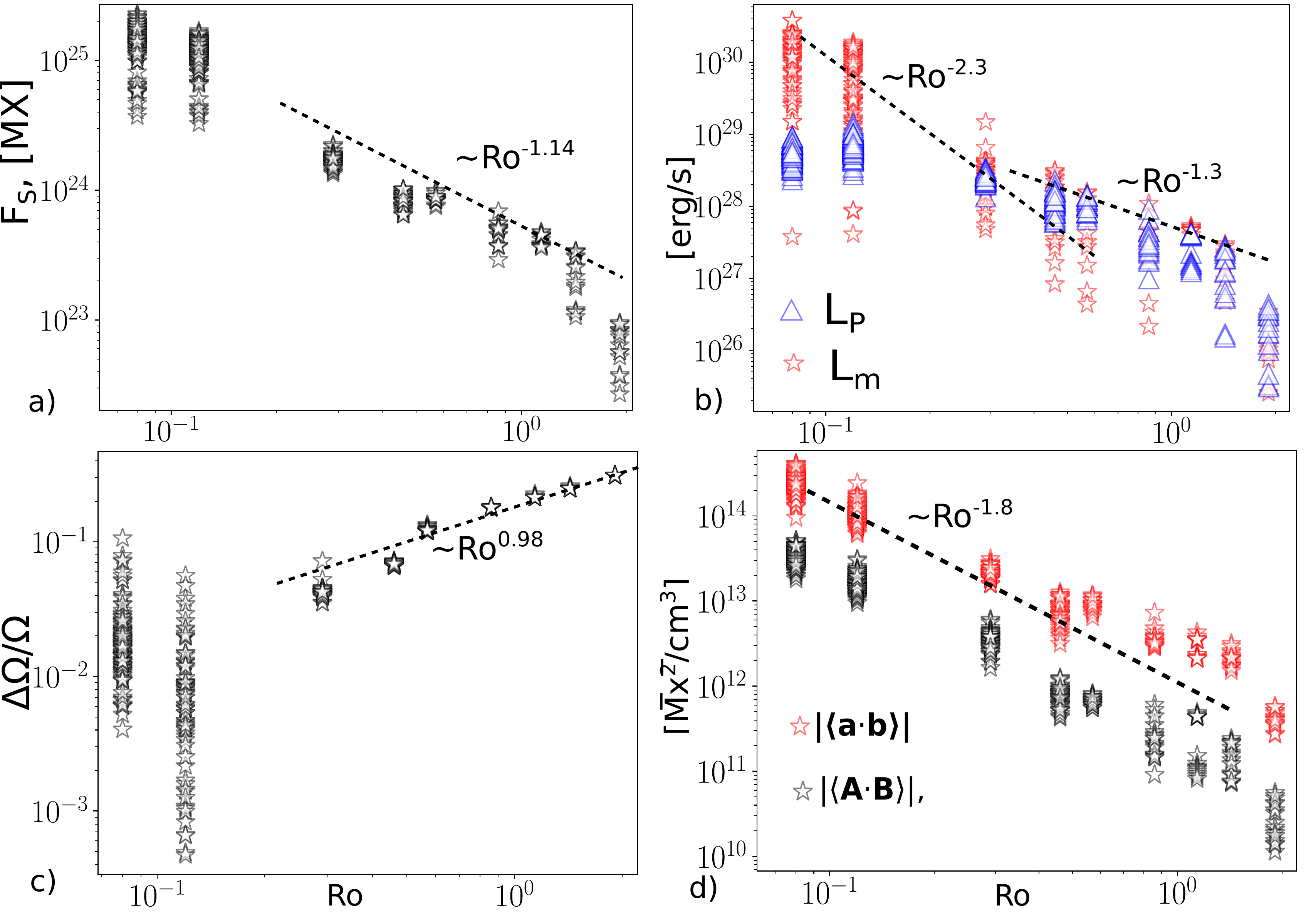}

\caption{\label{act-ros}a) {The magnitude of the magnetic flux generated
in the upper part of the convection zone and the Rossby number, where, each 
point represents results from a snapshot of the time series;} b) the
Poynting luminosity variations (blue triangles) and the irradiance
variations (red triangles) vs the Rossby number; c) the magnitude
of the latitudinal shear at the surface vs the Rossby number;
d) the same as c) for the magnitude of the mean large-scale magnetic
helicity density at the surface.}
\end{figure}

The stellar rotation - magnetic activity relations are often considered
as a major argument in favor of the turbulent dynamo action in convection
zones of the late-type stars \citep{Noyes1984,Baliunas1995}. Figure
\ref{act-ros}a) shows the dependence of the magnitude of the total
magnetic flux generated in the convection zone on the Rossby number.
We find that the studied interval of rotation rates includes both
saturation regimes. The star with a rotation period of 30 days shows
a considerable drop in the generated magnetic flux in comparison with
the solar case. At the opposite end, for the case of the small Rossby
number we see a sign of plateau, which is characterized by an increase
in the magnetic energy variability. This result is in agreement with
observations of \cite{Noyes1984,Vidotto2014,See2015}. {The
power-law for the generated flux vs the Rossby number is in agreement
with results of \cite{Vidotto2014}, who found the power-law $\mathrm{\Phi_{V}\sim Ro^{-1.19}}$,
where $\Phi_{V}$ is the total magnetic flux at the surface. We assume,
that in the solar-type stars the surface magnetic flux is originated
due to the magnetic active regions emergence, and the source of the
magnetic flux is provided by the toroidal magnetic field in the stellar
interior. Therefore, we assume that $\mathrm{\Phi_{V}\sim F_{S}}.$
The models show $F_{S}\sim\mathrm{Ro^{-1.14}}$, which is in agreement
with observations. The models show that the Poynting flux has power-law
$L_{P}\sim\mathrm{Ro^{-1.3}}$, then we have $L_{P}\sim\mathrm{F_{S}^{1.1}}$.
This scaling is in agreement with results of \cite{Pevtsov2003a}
and \cite{Vidotto2014} for the X-ray luminosity of the magneto-active
stars. }

Our simulations touch the low boundary of the dynamo saturation regime
at the small Rossby number. Similarly, such a plateau is found for
the magnitudes of the Poynting flux and the variations of the heat
flux. The Poynting flux describes the magnetic energy input in the
outer atmosphere of the star. Following \cite{Kleeorin1995,Blackman2015},
the magnitude of the flux can serve as an estimation of the X-ray radiation
of the stellar corona. \cite{Wright2016} found the saturation of
the X-ray luminosity occurs at Ro\textasciitilde 0.1. This roughly
agrees with the results shown in Figure \ref{act-ros}b). Still, our
conclusions are preliminary because we have only two models that are
in the saturation regime. These models show that the magnitude of
the relative latitudinal shear varies with STD level $\sigma\approx1$
(Figure \ref{act-ros}c)). The branch of models below the saturation
regime shows relation $\Delta\Omega/\Omega\sim\mathrm{Ro^{-0.98}}$,
which is in agreement with observations \citep{Barnes2005,Saar2011}
and expectations of the linear mean-field theory \citep{Kitchatinov1999a}.
\begin{figure}
\includegraphics[width=0.99\columnwidth]{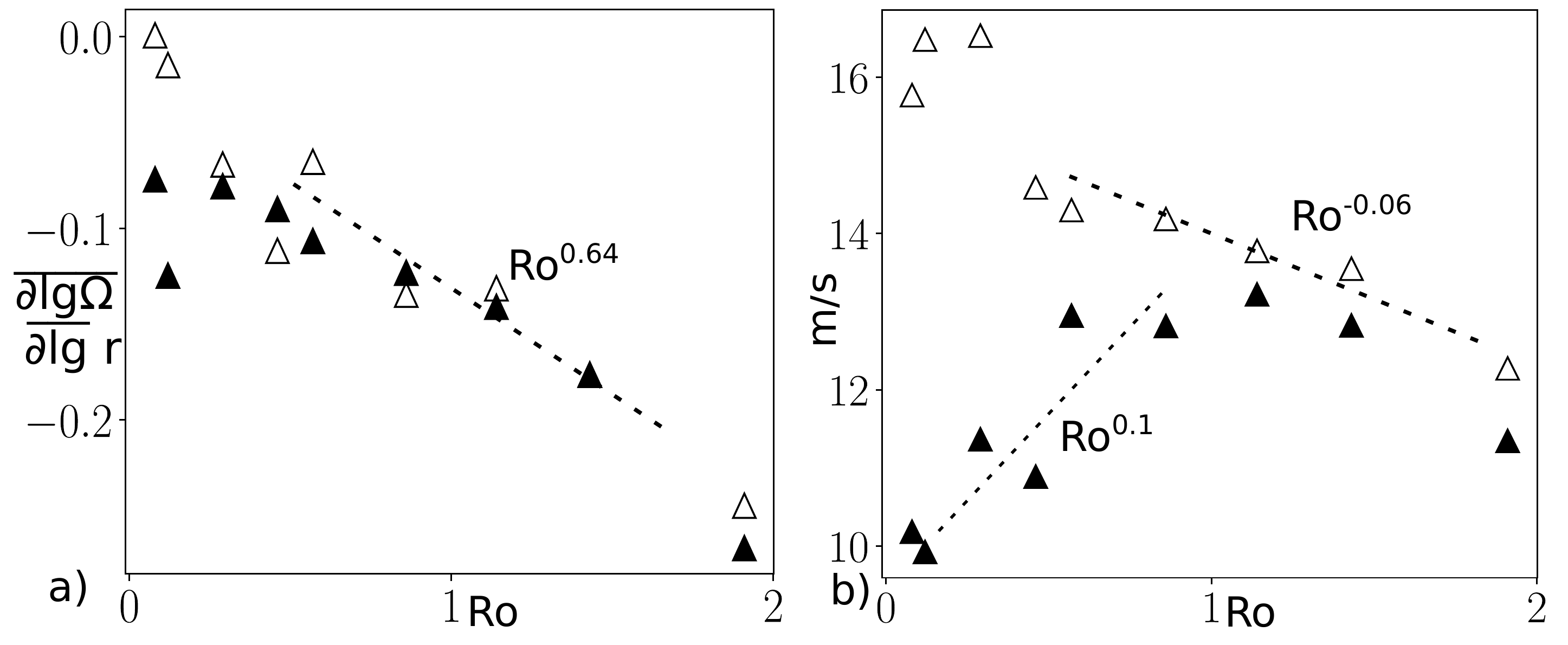}

\caption{\label{ros-mc}{a)The mean radial shear in the bulk of the
convection zone vs the Rossby number; the white triangles show the
kinematic dynamo models and black triangles stand for the non-kinematic
dynamo models. b)The magnitude of the meridional circulation at the
surface vs the stellar Rossby number. }}
\end{figure}

{Figure \ref{ros-mc} shows the relations of the mean magnitude
of the radial shear and the magnitude of the surface meridional circulation
with the stellar Rossby number. The mean radial shear in the bulk
of the convection zone is negative for all the models except the model
M1 (kinematic). In the kinematic dynamo models, the absolute magnitude
of shear does not follow the unique power-law showing the exponential
decrease with the decrease of the rotational period. This is in a qualitative
agreement with the results of \cite{Kitchatinov1999a}. For the range
of periods between 8 and 30 days the relationship is close to the
power-law $\mathrm{Ro^{0.64}}.$ The non-kinematic models show the
stronger radial shear for the fast rotation case. The surface meridional
circulation is directed to the pole in all cases. The fast-rotating
stars show the strong concentration of the circulation speed toward
the surface. The non-kinematic dynamo models of the fast rotating
stars show a shallow inversion of the circulation speed as well (see,
Figure \ref{fig:cvrms}). The kinematic dynamo models show a
slight increase of the circulation speed with a decrease of the rotational
period in following to the power-law $\mathrm{Ro^{-0.06}}$. The increase
qualitatively follows to results of \cite{Kitchatinov2011b}. Though
it is much smaller in our case because the meridional circulation
is quenched near the bottom of the convection zone. The nonkinematic
dynamo models show the non-monotonic relationship. }
\begin{figure}
\includegraphics[width=0.99\columnwidth]{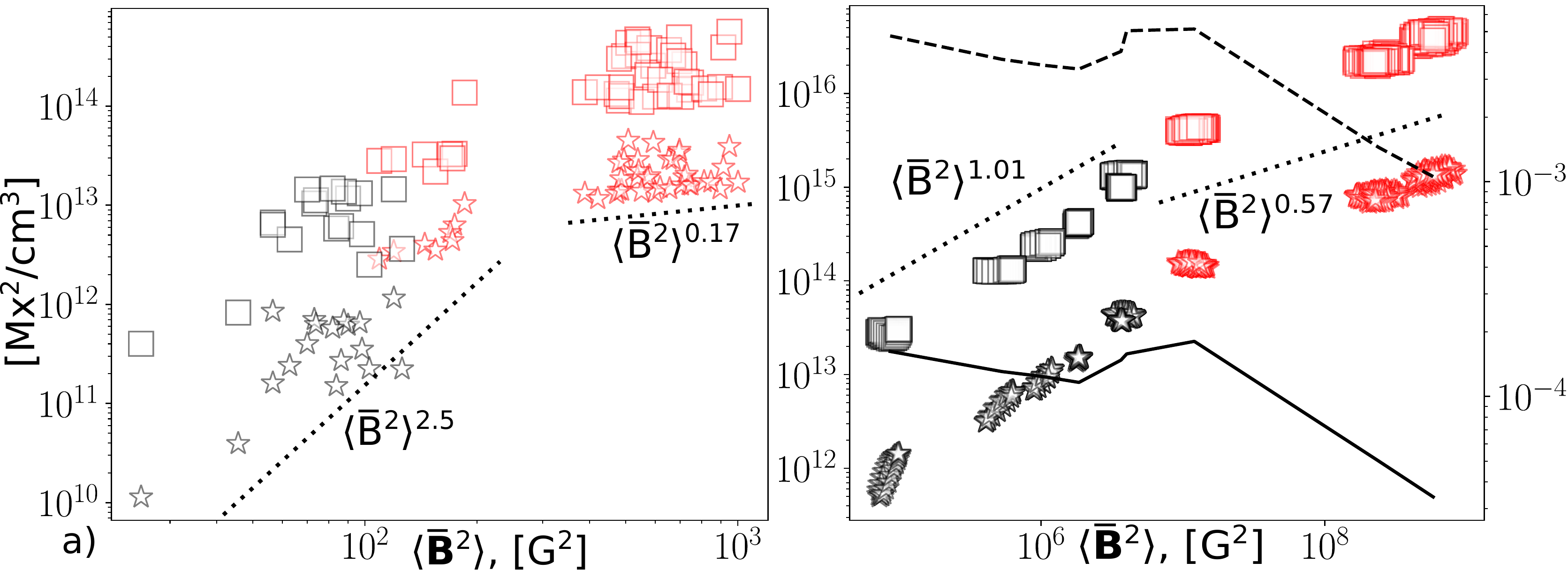}

\caption{\label{helab}a) Relation of the large- (stars) and small-scale (squares)
helicity density with the mean square density of the surface large-scale
magnetic field, the red color stars show samples from the runs M1n,
M2n and M5n; b ) points show the same for the mean parameters in bulk
of the convection zone; the solid line shows the ratio $\left|\left\langle \boldsymbol{\overline{A}}\cdot\boldsymbol{\overline{B}}\right\rangle \right|/\left(R\left\langle \boldsymbol{\overline{B}^{2}}\right\rangle \right)$
and the dashed line shows the same for the total magnetic helicity
density, $\left(\left|\left\langle \boldsymbol{\overline{A}}\cdot\boldsymbol{\overline{B}}\right\rangle \right|+\left|\left\langle \mathbf{a}\cdot\mathbf{b}\right\rangle \right|\right)/\left(R\left\langle \boldsymbol{\overline{B}^{2}}\right\rangle \right)$.}
\end{figure}

{Figure \ref{act-ros}d shows relations between the magnitudes
of the large- and small-scale magnetic helicity density and the Rossby
number. The small-scale helicity density is an order of magnitude
larger than the large-scale helicity density. For the large-scale
helicity our results agree with the recent survey of \cite{Lund2020}.
Similarly to the above-cited paper, we looked at relations of the
mean energy of the large-scale magnetic fields at the surface with
the magnitude of the mean large-scale helicity density, $\left|\left\langle \boldsymbol{\overline{A}}\cdot\boldsymbol{\overline{B}}\right\rangle \right|\propto\left\langle \boldsymbol{\overline{B}^{2}}\right\rangle ^{\alpha}$.
They are shown in Figure \ref{helab}a. We find that the fast and
slow rotating stars show the different scaling laws. In the stars
with the rotational period less than 5 days the generated large-scale
magnetic helicity density depends weakly on the mean magnetic energy
density, showing the power-law $\alpha\approx0.17$. The branch of
stars with the rotational period higher than 5 days show power-law
with $\alpha\approx2.5$. We find similar relations for the mean
small-scale helicity density. According to the results of \cite{Lund2020}
we analyzed the scaling laws of the mean helicity density for the
energy of the toroidal and poloidal component of the magnetic field separately.
For the toroidal magnetic field, we get qualitatively the same results
as shown in Figure \ref{helab}. This is because in our models the
mean strength of the surface toroidal magnetic is about of factor
3 larger than the mean strength of the surface poloidal magnetic field.
For the poloidal magnetic field we get the unique power law for all
runs, $\left|\left\langle \boldsymbol{\overline{A}}\cdot\boldsymbol{\overline{B}}\right\rangle \right|\propto\left\langle \boldsymbol{\overline{B}_{P}^{2}}\right\rangle ^{2.5}$.
Note, that a considerable set of stars from study of \cite{Lund2020}
are stars with a mass low than the Sun and they can operate another
type of  large-scale dynamo, which is due to dynamo instability
of the large-scale nonaxisymmetric magnetic field. Our study confirms
the conclusion of \cite{Lund2020} that different scaling laws in
the set of magneto-active stars may indicate the different dynamo
regimes.}

{In general, for the mean helicity density in the volume we
expect (\cite{Moffatt1978,Arnold1992}): 
\begin{equation}
\left|\left\langle \boldsymbol{\overline{A}}\cdot\boldsymbol{\overline{B}}\right\rangle \right|<C\left\langle \boldsymbol{\overline{B}^{2}}\right\rangle \label{eq:ab2}
\end{equation}
where $C\sim R$ is a positive dimension constant. Results in Figure\ref{helab}b
shows the volume-averaged parameters of the magnetic helicity density
and the mean square density of the large-scale magnetic field inside
the dynamo region. We find that the decrease of the rotational periods
from 30 to 8 days results in an increase of the generated large-scale
magnetic helicity proportionally to the energy of the large-scale
magnetic field. The efficiency of the magnetic helicity production
is reduced for the stars with the rotational periods less than 5 days.
Also, we find that the mean twist of the large-scale magnetic field
in the bulk of the convection zone, i.e., the ratio }$\left|\left\langle \boldsymbol{\overline{A}}\cdot\boldsymbol{\overline{B}}\right\rangle \right|/\left(R\left\langle \boldsymbol{\overline{B}^{2}}\right\rangle \right)${,
is reduced with the increase of the rotation rate. }

\begin{figure*}
\includegraphics[width=0.95\textwidth]{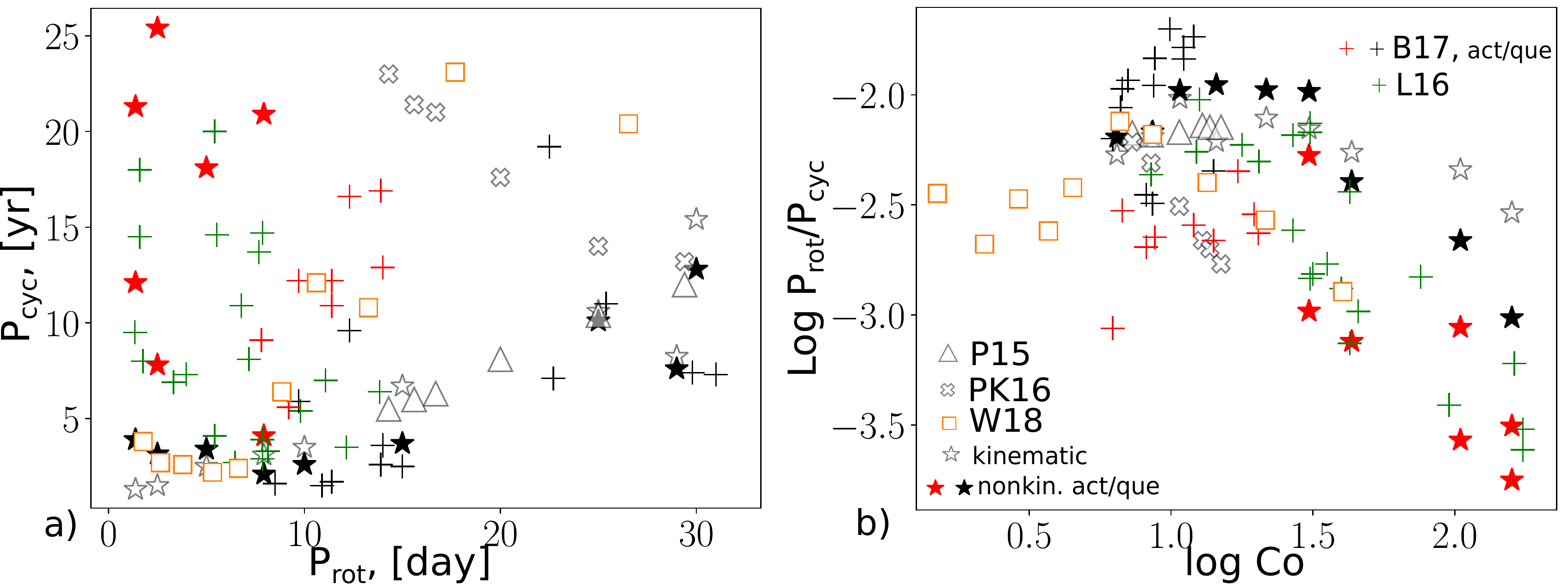}

\caption{\label{perper}a) Relation of the dynamo period with stellar rotation
period: the red and black crosses include results of Brandenburg et
al (2017) (B17) for the F and G-stars, the red crosses mark the  "active"
branch and the red crosses mark the quiet branch; the green crosses
mark results of Lehtinen et al(2016)(L16) for the young solar-type
stars; the white triangles show the quasi-non-kinematic dynamo models
of Pipin(2015) (P15), the white crosses show the results of the kinematic
dynamo model Pipin \&Kosovichev(2016) (PK16) for the model without
magnetic buoyancy effect; the orange squares show the results of Warnecke(2018)
(W18), the white stars show the kinematic models of this paper and
the black and red stars show the non-kinematic models, where the red
stars mark the long dynamo period. Panel b) shows the dependence on
the Coriolis number, where $\mathrm{Co}=4\pi/\mathrm{Ro}$.}
\end{figure*}

We employ wavelet analysis to identify the dynamo periods. For
the time series, we use both the time-latitude diagrams of the toroidal
magnetic field and the radiation flux variations. We also studied
the time-series of the integral parameters including the total magnetic
flux, the Poynting flux luminosity, and the total irradiance variation.
For the rotation periods longer than 10 days all parameters give a
unique value for the dynamo period that corresponds to the period
of the near-surface dynamo waves of the large-scale toroidal magnetic
field. For the sample of the fast rotating stars, the time-latitude
diagrams of the toroidal magnetic field give the most accurate estimation
of the main dynamo period. The series of the Poynting flux luminosity
and irradiance variations can give twice shorter dynamo periods, (cf.,
e.g. Fig\ref{fig:nkinl-1} and \ref{fig:nkinl})). Also, the series
of models for rotational periods of 8 and fewer days are non-stationary.
There we find the long periods in their dynamo patterns. 

Figure\ref{perper} compares results of a sample of F- and G-stars
of \cite{Brandenburg2017A}, the young solar-type stars of \cite{Lehtinen2016},
and different dynamo models for a relation of the dynamo cycle period
on stellar rotation period and the reverse Rossby number. The latter
is also called the Coriolis number. The earlier results of \cite{Saar1999}
and \cite{Bohm-Vitense2007} suggested that the decrease of the dynamo
period with the increase of the rotation rate can have different branches
This phenomenon is attributed to the ``active'' and ``quiet'' states of
the stellar magnetic activity. Their results were further refined
by \cite{Brandenburg2017A} and other studies \cite{Olah2009} and
\cite{Olspert2018}. The latter two and the survey of \cite{Lehtinen2016}
did not show a clear division on the quiet and active branches.
The set of the dynamo models which we use for comparison with observations
includes the quasi-non-kinematic dynamo models of \cite{Pipin2015},
one kinematic dynamo model from results of \cite{Pipin2016b}, the
results of global convection simulations of solar-like stars of \cite{Warnecke2018A}
(we remove one extreme case from his set) and sets of the kinematic
and non-kinematic models of this paper. For the rotation period interval
of 10 to 30 days, our models (both versions) show the linear increase
of the dynamo period with an increase of the rotational period. The
slope of the linear expression is about 0.36$\pm0.03$. Expressing
the dynamo period in days we find the linear slope is 136$\pm13$
which is about factor 4 higher than the results of \cite{Olah2009}.
We find that for the linear branch of models in the interval from
5 to 30 days the dynamo period depends on the amplitude of the magnetic
buoyancy effect. In our models, we include the default expression
of the mean-field magnetic buoyancy which follows from the results
of \cite{Kitchatinov1993}. \cite{Pipin2016b} studied the different
types of kinematic dynamo models and found that neglecting the magnetic
buoyancy results in inversion of the slope sign (see the set of the
white crosses in Figure \ref{perper}). Such a slope is typical for
the results of the flux-transport models \cite{Jouve2010}. The results
of \cite{Warnecke2018A} show about factor 2 larger slope than ours.
His results fit well into the ``active'' branch of stars from a
sample of \cite{Brandenburg2017A}. Our models for the period of rotation
from 10 to 30 days fit approximately into the ``quiet'' branch of
the sample from \cite{Brandenburg2017A}.

For the stars rotating with a period less than 10 days, we find a
weak dependence of the main dynamo period on the stellar rotation
rate. Similarly to \cite{Warnecke2018A}, it is found that the decrease
of the rotation period from 5 to 1 day results in a variation in the
dynamo period from about 3 to 4 years. The models M1dn and M2dn show
the non-stationary variations of the magnetic cycle parameters with
strong cycle-to-cycle parity variations. In these models, we find
the cyclic patterns with periods that are much longer than the main
period determined by the near-surface dynamo waves. We find the long
dynamo periods for about 12 and 16 years for models M1dn and M2dn
respectively. In Figure \ref{perper}a, these points are in the cloud
of the young solar-type stars sample of \cite{Lehtinen2016}.

Normalization of the dynamo cycle frequency to the stellar rotation 
rate was found to be a good parameter to distinguish between the ``active''
and ``quiet'' branches of the magnetic activity \citep{Saar1999}.
Figure\ref{perper} b show this parameter against the reverse Rossby
number. Indeed, the sample of F- and G-stars of \cite{Brandenburg2017A}
shows the distinct branches of stellar activity, where the ``active''
branch has longer periods. Similarly to \cite{Warnecke2018A}, the
set of our models fits well the sample of \cite{Lehtinen2016}. \cite{Viviani2018}
found similar dependence in global convection simulations non-axisymmetric
dynamo models. We see, that the branch of models, where the dynamo
period decreases with the increase of the rotation rate (rotation
periods between 30-10 days), is slightly inclined from the direction
of the abscissa axis. The slope is opposite to that in the samples
of \cite{Brandenburg2017A} and \cite{Warnecke2018A}. This branch
can be identified in the kinematic dynamo models or in the full nonlinear
models when the effect of the magnetic field on the large-scale flow
is not strong. The models M8dn M5dn, M2dn, and M1dn follows to results
\cite{Lehtinen2016} both for the ``short'' and long dynamo periods.
Note, that the slope of this branch agrees approximately with all
models that show the increase of the dynamo period with the increase
of the rotation rate, e.g., the sample of models from \cite{Pipin2016b}
or results of \cite{Strugarek2017} (see, \citealp{Warnecke2018A}).
In the whole, the results presented in Figure\ref{perper} are compatible
with the surveys of \cite{Olspert2018}, \cite{BoroSaikia2018} and
\cite{Lehtinen2020}.

\section{Discussion}

In the paper, we explore the magnetic cycle parameters for the solar-type
star with a rotation period from 1 to 30 days. For this study, we
employ the non-kinematic mean-field dynamo models which take into
account the effect of the magnetic activity on the angular momentum
and heat transport inside of the convection zone. The given model
is fully compatible with the solar dynamo model suggested recently
by \cite{Pipin2019c} as the model of the solar torsional oscillations.
In our previous study, \citep{Pipin2015,Pipin2016b} we used a smaller
interval of the rotational periods. Also, that studies did not take
into account the effects of the meridional circulation neither in
the advection of the large-scale magnetic field nor in the angular
momentum and heat transport in balk of the convection zone. Despite
the difference, both the previous and the current studies consider
the dynamo process distributed in the whole convection zone. In all
the models the near-surface dynamo wave propagation patterns satisfy
the Parker-Yoshimura rule \cite{Yoshimura1975}. We find that the
efficiency of the large-scale dynamo grows with the increase of the
rotation rate. The equipartition strength of the large-scale magnetic
field for the star rotating with a period of 2 days is twice as high
as the solar analog rotating with a period of 25 days.

In the kinematic models, we find that the decrease of the rotation
below 10 days results in the increasing complexity of the near-surface
dynamo wave pattern. Near the pole we find the wave propagating toward
the equator. Following the Parker-Yoshimura rule, this is because
of the nearly radial angular velocity profiles and the positive (at
the North) $\alpha$--effect in the main part of the convection zone.
The cylinder-like angular velocity profiles near the equator cause
the poleward propagation of the dynamo waves. The dynamo period of
the kinematic models increases with the decrease of the rotation rate.
This is compatible with the results of studies exploring the chromospheric
and photometric variations of the solar-type stars \citep{Olah2009,Olspert2018,BoroSaikia2018}.
Similar results are suggested by the global simulations of \cite{Guerrero2019}.
In our previous study, we find that this relation can depend on details
of the dynamo models. For example, the distributed dynamo model with
the standard angular velocity and $\alpha$-effect profiles (see,
\cite{Pipin2016b}) can show the inverse ``period-period'' relation
in case if the magnetic buoyancy effect is disregarded. Similar dependence
was found in flux-transport models of \cite{Jouve2010} and global
simulations of \cite{Strugarek2017}. The kinematic dynamo models
do not show a clear sign of the magnetic activity saturation with
an increase in rotation rate.

The non-kinematic models show that saturation of the magnetic activity
is likely happening for the rotation period of less than 5 days. In general,
this conclusion is in agreement with observations, e.g., \cite{Wright2016}.
Note, that the partially- and fully-convective stars can show a difference
in parameters of the magnetic activity saturation \citep{Nizamov2017}.
Therefore the validity of our conclusion can be questioned for the
general case of the non-axisymmetric dynamo, which becomes dominant
for the fast-rotating solar-type stars \citep{Viviani2018}. The non-kinematic
models in the range of rotation periods longer than 15 days agree
qualitatively with their kinematic analogs. In the model with a rotation
period of 15 days, we find the doubling of the magnetic cycle frequency.
This model is likely to show the solution for the marginal state.
The marginal state is caused by the re-organization of the large-scale
flow which shows the multiple meridional circulation cell in the bulk
of the convection zone. Interestingly, the model M15n shows the mixed
parity solution with the approximate energy equipartition of the symmetric
and antisymmetric about the solar equator toroidal magnetic field.
Such a type of dynamo solution is often considered to be typical for
the solar Grand minima events \citep{Sokoloff1994,Weiss2016}. Interesting
that the dynamo models M10n and M15n shows the reduction of the relative
variations of the integral magnetic parameters $F_{T}$ and $F_{S}$
in comparison with the kinematic cases and with the results of the
models M20n and M25n. Using only the integral traces of the magnetic
activity, the real star with such type dynamo evolution is likely
to be identified as a star in a Maunder minimum state \citep{Baliunas1995}.
The mean unsigned flux of the large-scale toroidal magnetic field
in the model M10n is about 2$\cdot10^{24}$ Mx which higher than mean
$F_{T}$ in the models M15n and M25n. \cite{Schrijver1984} found
that the total magnetic flux observed in the solar cycle can be an
order of $10^{24}.$ This flux includes both the large- and small
scales magnetic fields.

The non-kinematic models show a transition from the state with one
cell per hemisphere to the state with the multiple meridional circulation
cells. Note, that the number of the meridional cells in the solar
convection zone is still under debate \citep{Zhao2013,Rajaguru2015,Boning2017}.
Here we deliberately use the reference model with one meridional cell
per hemisphere. The break of the global circulation cell causes a
number of consequences for the dynamo wave propagation and evolution
of the radial magnetic field on the surface. For example, in the model
M10n, it causes the weakening of the toroidal magnetic field near
the equator. Indeed, the toroidal magnetic field dynamo waves propagating
from the bottom of the convection zone to the top are trapped near
saddle-type stationary point of the meridional circulation, and the
cylinder-like angular velocity distribution blocks the dynamo wave
propagation toward the equator. For the fast rotating case, the differential
rotation is suppressed and the direction of the meridional circulation
in the main part of the convection zone is reversed in comparison
to the solar case. This provides the equator-ward propagation of the
toroidal magnetic field in the models M1n and M2n. Also, the dynamo
period in these models is longer and the latitudinal scale of the
toroidal magnetic field is larger than in the models M5n, M8n, and
M10n. {We find that the transition of the dynamo with one circulation
cells to the case with the multiple circulation cells results into
reduction the efficiency of the large-scale magnetic helicity production.
This is due to both the increase of the toroidal magnetic field concentration
toward the surface and the high ratio $B_{T}/B_{P}$ in the models
M1n, M2n and M5n. In our models we neglect the nonaxisymmetric component
of the large-scale dynamo. Combing our results, the results of the
direct numerical simulation of \cite{Viviani2018} and the results
of stellar magnetic activity observations of \cite{Lund2020} we can
conjecture that for the fast rotating stars the large-scale magnetic
helicity production can be due to the nonaxisymmetric dynamo. }

The kinematic models show the monotonic decrease of the dynamo period
with the increase of the rotation rate. The non-kinematic models show
the decrease of the dynamo period in the range of the rotational periods
from 30 to 15 days. This decrease is not as strong as found for the
``quiet'' branch of F- and G- stars by \cite{Brandenburg2017A}.
Therefore the models show the inverse inclination on the diagram of
Figure \ref{perper}b. Our previous results, \cite{Pipin2015}, for
the distributed dynamo model without the meridional circulation show
a better agreement with \cite{Brandenburg2017A}. Note , that the
extended analysis of the chromospheric activity in cool stars by \cite{BoroSaikia2018}
shows no clear distinction between active and quiet branches. For
the fast rotating case, the non-kinematic models, in contrast with
their kinematic analogs, show no trend in the primary dynamo period
with an increase of the rotation rate. The global convection simulations
of \cite{Warnecke2018A} do not show clear trend in this case as well.
The primary and secondary periods in the models for the range of the
rotational periods from 1 to 8 days agrees with the findings from
observations of the fast rotating solar analogs \citep{Lehtinen2016}.

We find that a study of the dynamo periods solely on the base of the
integral proxies of stellar magnetic activity can bias conclusions
about the magnetic periodicity of the solar-type stars. Using our
results we can identify several traps of such studies. Firstly, the
mix of the magnetic parity modes can result in variations of the integral
parameters showing either a state of ``Maunder minimum'' or the
double dynamo frequency oscillations. Secondly, the strong nonlinearity
of the dynamo solution can cause variations of the global activity
parameters with the double frequency (\citealp{Setal20}), as well.
In the solar case, the effect is not strong. However, it can bias
the conclusion for the fast-rotating solar analogs, which are expected
to show a highly nonlinear dynamo regime (cf., $\beta$ -parameter
in Tables\ref{tab} and \ref{tab-2}). Finally, in the nonlinear dynamo
regimes, the integral parameters show the non-stationary evolution
where the main period of the time-latitude dynamo waves is hard to
identify as the primary dynamo period. The models M1n and M2n show
good examples of this type.

In our study, we investigated a few aspects of the observational trends
of the magnetic variability of the solar-type stars. Our discussion
has been concentrated on the properties of the near-surface dynamo
wave patterns and their relation with the dynamics of the large-scale
flow and variations of the integral proxies of the magnetic activity.
We did not touch numerous theoretical aspects of the magnetic activity
of the solar-type stars including changes in the magnetic field topology
and the type of axial symmetry of the large-scale magnetic field with
the rotation rate of a star. These and other interesting questions
can be studied further using the numerical simulations and the growing
base of stellar magnetic activity observations.

\textbf{Data Availability Statements.} The data of the kinematic and
non- kinematic models together with the PYTHON scripts are available
at \href{https://drive.google.com/drive/folders/10qFkZiAz_U6tZFZsq7flsIiC0RzO39cIhttps://drive.google.com/drive/folders/10qFkZiAz_U6tZFZsq7flsIiC0RzO39cI}{google-drive}

\textbf{Acknowledgements} The author acknowledge the financial support
by the Russian Foundation for Basic Research grant 19-52-53045 and
support of scientific project FR II.16 of ISTP SB RAS. 

 \bibliographystyle{mnras}
\bibliography{dyn}

\end{document}